\documentclass[12pt]{article}

\usepackage{xcolor}
\usepackage{graphicx}
\usepackage{float}
\usepackage{cite}
\usepackage{amsfonts}
\usepackage{amssymb}
\usepackage{amsmath}
\usepackage{relsize}
\usepackage[compatibility=false]{caption}
\usepackage{subcaption}

\newcommand{\lc}{\left\{}
\newcommand{\rc}{\right\}}
\newcommand{\ls}{\left[}
\newcommand{\rs}{\right]}

\def\gtwid{\mathrel{\raise.3ex\hbox{$>$\kern-.75em\lower1ex\hbox{$\sim$}}}}
\def\ltwid{\mathrel{\raise.3ex\hbox{$<$\kern-.75em\lower1ex\hbox{$\sim$}}}}
\def\square{\kern1pt\vbox{\hrule height 1.2pt\hbox{\vrule width 1.2pt\hskip 3pt
   \vbox{\vskip 6pt}\hskip 3pt\vrule width 0.6pt}\hrule height 0.6pt}\kern1pt}

\begin{document}

\begin{titlepage}

\begin{flushright}
UFIFT-QG-20-04
\end{flushright}

\vskip 2cm

\begin{center}
{\bf Inflaton Effective Potential from Fermions for General $\epsilon$}
\end{center}

\vskip 1cm

\begin{center}
A. Sivasankaran$^{*}$ and R. P. Woodard$^{\dagger}$
\end{center}

\vskip .5cm

\begin{center}
\it{Department of Physics, University of Florida,\\
Gainesville, FL 32611, UNITED STATES}
\end{center}

\vspace{1cm}

\begin{center}
ABSTRACT
\end{center}
We accurately approximate the contribution of a Yukawa-coupled fermion 
to the inflaton effective potential for inflationary geometries with a
general first slow roll parameter $\epsilon(t)$. For $\epsilon = 0$ 
our final result agrees with the famous computation of Candelas and 
Raine done long ago on de Sitter background \cite{Candelas:1975du}, 
and both computations degenerate to the result of Coleman and Weinberg 
in the flat space limit \cite{Coleman:1973jx}. Our result contains a 
small part that depends nonlocally on the inflationary geometry. Even 
in the numerically larger local part, very little of the $\epsilon$ 
dependence takes the form of Ricci scalars. We discuss the implications 
of these corrections for inflation.

\begin{flushleft}
\end{flushleft}

\vskip .5cm

\begin{flushleft}
$^{*}$ e-mail: aneeshs@ufl.edu \\
$^{\dagger}$ e-mail: woodard@phys.ufl.edu
\end{flushleft}

\end{titlepage}

\section{Introduction}

The most recent data on primordial perturbations \cite{Aghanim:2018eyx} are
consistent with the simplest models of inflation based on gravity plus a single, 
minimally coupled scalar inflaton $\varphi$,
\begin{equation}
\mathcal{L}_{\rm inflaton} = \frac{R \sqrt{-g}}{16 \pi G} -\frac12 \partial_{\mu} 
\varphi \partial_{\nu} \varphi g^{\mu\nu} \sqrt{-g} - V(\varphi) \sqrt{-g} \; . 
\label{Linflaton}
\end{equation}
Once inflation is established the system rapidly approaches homogeneity and 
isotropy, which means $\varphi \rightarrow \varphi_0(t)$ and,
\begin{equation}
ds^2 = -dt^2 + a^2(t) d\vec{x} \!\cdot\! d\vec{x} \qquad \Longrightarrow
\qquad H(t) \equiv \frac{\dot{a}}{a} \;\; , \;\; \epsilon(t) \equiv -
\frac{\dot{H}}{H^2} \; . \label{geometry}
\end{equation}
Inflation proceeds as long as the inflaton's potential energy dominates the
nontrivial Einstein equations,
\begin{eqnarray}
3 H^2 & = & 8\pi G \Bigl[ \frac12 \dot{\varphi}_0^2 + V(\varphi_0) \Bigr] \; ,
\label{E1} \\
- (3 \!-\! 2 \epsilon) H^2 & = & 8 \pi G \Bigl[ \frac12 \dot{\varphi}^2_0
- V(\varphi_0) \Bigr] \; , \label{E2}
\end{eqnarray}
Hubble friction slows the inflaton's roll down its potential,
\begin{equation}
\ddot{\varphi}_0 + 3 H \dot{\varphi}_0 + V'(\varphi_0) = 0 \qquad \Longrightarrow
\qquad \dot{\varphi}_0 \simeq -\frac1{\sqrt{24 \pi G}} \frac{V'(\varphi_0)}{
\sqrt{V(\varphi_0)}} \; . \label{scalareqn}
\end{equation}
At the end of inflation the scalar's potential energy falls to become 
comparable to its kinetic energy, which reduces Hubble friction and allows
$\varphi$ to rapidly oscillate around the minimum of its potential. During 
this phase of ``reheating'' the inflaton's kinetic energy is transferred 
into a hot, dense plasma of ordinary particles and Big Bang cosmology follows
its usual course. 

Facilitating the transfer of inflaton kinetic energy into ordinary matter
during reheating obviously requires a coupling between $\varphi$ and ordinary 
matter. The one we shall study here is to a massless fermion,
\begin{equation}
\mathcal{L}_{\rm fermion} = \overline{\psi} \gamma^b e^{\mu}_{~ b} 
\Bigl( \partial_{\mu} \!+\! \frac{i}2 A_{\mu cd} J^{cd} \Bigr) \psi
\sqrt{-g} - f \varphi \overline{\psi} \psi \sqrt{-g} \; . \label{Lfermion}
\end{equation}
where $e_{\mu c}(x)$ is the vierbein (with $g_{\mu\nu}(x) = e_{\mu a}(x) 
e_{\nu b}(x) \eta^{ab}$), $A_{\mu c d}(x) = e^{\nu}_{~c} (\partial_{\mu} 
e_{\nu d} - \Gamma^{\rho}_{~\mu\nu} e_{\rho d})$ is the spin connection, the 
$\gamma^a$ are gamma matrices (with $\{\gamma^a,\gamma^b\} = -2\eta^{ab} I$)
and $J^{cd} \equiv \frac{i}{4} [\gamma^c,\gamma^d]$. Such a coupling causes 
the 0-point energy of ordinary matter (in this case, the fermion) to induce 
corrections to the inflaton potential $V(\varphi)$ the same way that Coleman 
and Weinberg long ago demonstrated in flat space \cite{Coleman:1973jx},
\begin{equation}
\Delta V_{\rm flat}(\varphi) = -\frac{(f \varphi)^4}{8 \pi^2} \ln\Bigl( 
\frac{f \varphi}{s}\Bigr) \; . \label{flatDV}
\end{equation}
(Here $s$ is the renormalization scale.) The result on an inflationary 
background (\ref{geometry}) depends in a complicated way on $H$ and $\epsilon$ 
--- which we will elucidate --- but expression (\ref{flatDV}) is still the 
leading large field result \cite{Miao:2015oba}.

\begin{figure}[H]
\centering
\includegraphics[width=7cm]{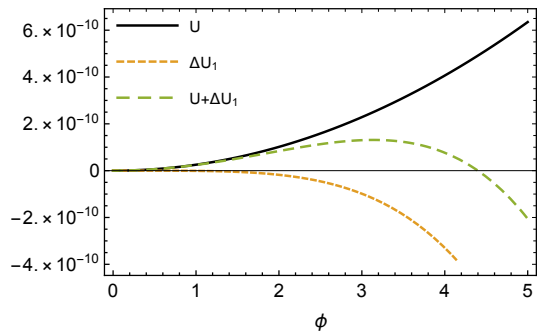}
\caption{\footnotesize{The solid black line is the dimensionless classical 
potential $U \equiv (8\pi G)^2 V = \frac12 c^2 \phi^2$ plotted as a function of the 
dimensionless inflaton field $\phi \equiv \sqrt{8\pi G} \times \varphi$. The dashed 
orange line is the one loop quantum correction for coupling $f^2 = 4.5\times 10^{-6}$,
and the dashed green line is the total potential. Note that solving the Horizon problem
in the classical model requires an initial value of $\phi_b \ltwid 14$, at which point
the total potential drives evolution towards a Big Rip singularity.}}
\label{CWplot}
\end{figure}

Cosmological Coleman-Weinberg potentials are potentially problematic for 
inflation because they can make significant changes to the classical 
trajectory of the inflaton \cite{Green:2007gs}. To fully explore the 
problem requires the dependence upon $H$ and $\epsilon$ that we shall 
determine, but the {\it possibility} of a problem is evident from the large 
field limiting form (\ref{flatDV}) which is plotted in Figure~\ref{CWplot}. 
Suppose the classical potential that drives inflation is the simple 
quadratic model, with its mass tuned to agree with the observed 
\cite{Aghanim:2018eyx} scalar amplitude,
\begin{equation}
V(\varphi) = \frac{c^2 \varphi^2}{16 \pi G} \qquad , \qquad c \simeq 7.1 \times
10^{-6} \; . \label{Vclass}
\end{equation}
\noindent Although the classical trajectory of $\varphi$ is towards zero starting 
from any initial value, the fact that fermionic corrections to the effective 
potential are {\it negative} means that a sufficiently large initial value 
drives the inflaton towards infinity, and a Big Rip singularity. Even if the 
coupling and the initial value are chosen to avoid this, the problem of fine 
tuning initial conditions has undergone a radical change from the classical 
model --- in which the kinetic, gradient and potential energy contributions 
are each unbounded above --- to the quantum-corrected model --- in which the 
kinetic and gradient energies can still be arbitrarily large but the potential 
energy is bounded. We do not assert that viable models are impossible, but one
must obviously take account of Cosmological Coleman-Weinberg potentials.

To quantify the potential for problems, we express the quantum correction 
(\ref{flatDV}) as a factor times the classical potential (\ref{Vclass}),
\begin{equation}
\Delta V_{\rm flat}(\varphi) = -\Bigl( \frac{f^2}{2 \pi c}\Bigr)^2 \!\times\! 
8\pi G \varphi^2 \ln\Bigl( \frac{f\varphi}{s}\Bigr) \!\times\! V(\varphi) \; .
\label{proportion}
\end{equation}
To estimate the classical initial value of $\varphi$, recall that the slow roll 
approximation for the number of e-foldings from the beginning ($\varphi_b$) of 
inflation to its end ($\varphi_e$) is,
\begin{equation}
N \simeq 8 \pi G \!\! \int_{\varphi_e}^{\varphi_b} \!\!
\frac{V(\varphi) d\varphi}{V'(\varphi)} = 2\pi G \Bigl( \varphi^2_b - 
\varphi^2_e\Bigr) \; .
\end{equation}
Because $\varphi_b \gg \varphi_e$ and $N$ must be greater than 50 to solve the 
horizon problem, we know that $\varphi^2_b \gtwid 200/8\pi G$. Hence the 
proportionality factor is,
\begin{equation}
\Bigl( \frac{f^2}{2 \pi c}\Bigr)^2 \!\times\! 8\pi G \varphi^2 \ln\Bigl( 
\frac{f\varphi}{s}\Bigr) \ltwid \Bigl( \frac{5 f^2}{\pi c}\Bigr)^2 
\ln\Bigl( \frac{200 f^2}{8 \pi G s^2}\Bigr) \; . \label{propest}
\end{equation}
How large the logarithm is depends on the unknown renormalization scale $s$, but
the inflaton changes so much that we can safely ignore it to conclude that
making the quantum correction have the same initial magnitude (but 
opposite sign) as the classical potential requires,
\begin{equation}
f^2 \ltwid \frac{\pi c}{5} \simeq 4.5 \times 10^{-6} \; .
\end{equation}
One can see from Figure~\ref{CWplot} that even such a small coupling would still 
leave the starting point on the wrong side of the total potential so that 
evolution would be towards a Big Rip singularity. Making the quantum correction
negligible would require correspondingly smaller couplings, which makes reheating
inefficient and requires changing the shape of the potential after the point at 
which observable perturbations are generated \cite{Mielczarek:2010ag,Liao:2018sci}.
This is explained in the Appendix. Again, we do not assert that Cosmological 
Coleman-Weinberg potentials preclude the possibility of developing viable models,
just that they must be considered. It should also be mentioned that there is 
{\it no alternative} to an order one Yukawa coupling for the top quark in Higgs 
inflation \cite{Bezrukov:2007ep}.

If cosmological Coleman-Weinberg potentials depended only on the inflaton they
could simply be subtracted. When gravity is dynamical the basic model 
(\ref{Linflaton}) is not renormalizable, so few cosmologists would quibble over
subtracting $\Delta V_{\rm flat}(\varphi)$ from the classical action. However, 
explicit computations on de Sitter background \cite{Candelas:1975du,Miao:2006pn} 
reveal a much more complex structure made possible by the addition of the 
dimensional parameter $H$,
\begin{equation}
\Delta V_{\rm dS} = -\frac{H^4}{8 \pi^2} \!\times\! F\Bigl( \frac{f \varphi}{H}
\Bigr) \; . \label{dSform}
\end{equation}
Strong indirect arguments indicate that (\ref{dSform}) remains approximately 
valid for the more general geometry (\ref{geometry}) of inflation 
\cite{Miao:2015oba} --- and these arguments were recently confirmed for scalar
couplings \cite{Kyriazis:2019xgj}, as well as pinning down the complex dependence 
on $\epsilon(t)$. This is crucial because the most general permissible subtraction
consists of a local function of $\varphi$ and the Ricci scalar, $R = 6 (2 - 
\epsilon) H^2$ \cite{Woodard:2006nt}. It follows that cosmological Coleman-Weinberg 
potentials cannot be completely subtracted, and studies show that the remainder 
after the best partial subtraction still makes disastrous changes 
\cite{Liao:2018sci,Miao:2019bnq}.

Rather than trying to subtract cosmological Coleman-Weinberg potentials a more hopeful
strategy is to arrange cancellations between the negative fermionic contributions
and the positive bosonic contributions \cite{Miao:2020zeh}. These cancellations 
would be exact in flat space but they cannot be exact on the geometry 
(\ref{geometry}) of inflation because they are not exact on de Sitter 
\cite{Candelas:1975du,Allen:1983dg,Miao:2006pn,Prokopec:2007ak,Miao:2015oba}.
The viability of bose-fermi cancellation depends upon how good the cancellation
is for general $\epsilon(t)$. A good approximation has been obtained for the 
cosmological Coleman-Weinberg potential induced by a minimally coupled scalar 
\cite{Kyriazis:2019xgj}; it is our purpose here to do the same for the fermion 
(\ref{Lfermion}). 

 Our derivation begins with the standard expression for the derivative 
of the effective potential as the coincidence limit of a fermion propagator whose 
mass is induced by its Yukawa coupling to the inflaton. We obtain the fermion
propagator by differentiating suitable scalar propagators which can be written as
spatial Fourier mode sums. All these results are exact and valid for any geometry
of the form (\ref{geometry}). What we approximate is the scalar mode functions,
testing our approximations by explicit numerical evolution. We also prove that our
approximation is sufficient to completely capture the divergence of the coincidence
limit, which we regulate using dimensional regularization. After renormalization 
our approximation expresses the effective potential as a part that depends on the
instantaneous values of $H(t)$, $\epsilon(t)$ and also $\dot{\epsilon}(t)$, plus
a numerically smaller part that depends nonlocally on the past evolution of the
geometry. In addition to the explicit numerical comparisons, we check that our
result degenerates to the known forms for flat space and for de Sitter. We also 
derive expansions which are valid for large and small field strengths.

In section 2 we derive a good approximation for the coincidence limit of a 
massive fermion propagator. Section 3 applies this result to compute the effective 
potential from (\ref{Lfermion}). Because our approximation becomes exact in the 
ultraviolet we can fully renormalize the result. Section 4 presents out
conclusions.

\section{Coincident Fermi Propagator for General $\epsilon$}

The purpose of this section is to derive a good analytic approximation for the
coincident massive fermion propagator in the general inflationary background 
(\ref{geometry}), which we consider to possess $D-1$ spatial dimensions to
facilitate dimensional regularization. The section begins by representing the 
fermion propagator in terms of scalar propagators with various masses and
conformal couplings. Their coincidence limits are then expressed as spatial
Fourier mode sums. A dimensionless equation is derived for the logarithm of the
amplitude. Graphical evidence is presented that this quantity has two phases,
and accurate analytic approximations are derived for each phase.

\subsection{Fermion to Scalar Propagators}

At one loop order, the Yukawa coupled fermion (\ref{Lfermion}) induces an 
effective potential $\Delta V$ whose derivative with respect to $\varphi$ obeys,
\begin{equation}
\Delta V'(\varphi) = \delta \xi \varphi R + \frac16 \delta \lambda \varphi^3 - 
f i \Bigl[\mbox{}_i S_i\Bigr](x;x) \; . \label{DVfermion}
\end{equation}
where $i[\mbox{}_i S_j](x;x')$ is the propagator of a fermion with mass $m = f
\varphi$ and $\delta \xi$ and $\delta \lambda$ are the coefficients of the 
conformal and quartic counterterms. There is a simple relation between the 
massive fermion propagator in a general inflationary background (\ref{geometry}) 
and scalar propagators $i\Delta[\xi,M^2](x;x')$ with various conformal coupling 
$\xi$ and mass $M^2$. If we change to conformal time (i.e., $d\eta = dt/a$) this 
relation is \cite{Candelas:1975du,Miao:2006pn},
\begin{eqnarray}
\lefteqn{i \Bigl[\mbox{}_i S_j\Bigr](x;x') = \frac1{a^{\frac{D-1}2}}
\Bigl[ i \gamma^{\mu} \partial_{\mu} + a m I\Bigr]
\frac{ a^{\frac{D-1}2}}{\sqrt{a a'} } } \nonumber \\
& & \hspace{1.3cm} \times \lc i
\Delta[\xi_c,M^2_+](x;x') \Bigl( \frac{I \!+\! \gamma^0}2\Bigr)
+ i \Delta[\xi_c,M^2_-](x;x') \Bigl( \frac{I \!-\! \gamma^0}2\Bigr)
\rc , \qquad \label{fermionprop}
\end{eqnarray}
where $\xi_c = \frac14 (\frac{D-4}{D-1})$ and $M^2_{\pm} = f \varphi (f \varphi 
\mp i H)$. 

The scalar propagators in expression (\ref{fermionprop}) satisfy the 
Klein-Gordon equation with conformal coupling, 
\begin{eqnarray}
\ls \square -\xi_c R - M^2_{\pm} \rs i \Delta[\xi_c,M^2_{\pm}](x,x') = 
\frac{i\delta^D(x-x')}{\sqrt{-g}} \; ,
\end{eqnarray}
where $\sqrt{-g} \square \equiv \partial_{\mu} \sqrt{-g} \, g^{\mu\nu} 
\partial_{\nu}$ is the covariant scalar d'Alembertian. The spinor trace of
the coincident fermion propagator in expression (\ref{fermionprop}) is,
\begin{eqnarray}
\lefteqn{i\Bigl[\mbox{}_i S_i\Bigr](x;x) = 2m \Bigl\{ i\Delta[\xi_c,M^2_+](x;x) 
+ i \Delta[\xi_c,M^2_-](x;x) \Bigr\} } \nonumber \\ 
& & \hspace{1.5cm} + i \Bigl[ \frac{\partial}{\partial t} + (D \!-\!2) H \Bigr] 
\Bigl\{ i \Delta[\xi_c,M^2_+](x;x) - i \Delta[\xi_c,M^2_-](x;x) \Bigr\} \; .
\qquad \label{fermiontrace}
\end{eqnarray}
To reach this form we have used,
\begin{equation}
\lim_{x \to x'} \gamma^{\mu} \partial_{\mu} i \Delta[\xi,M^2_{\pm}](x;x') = 
\frac12 \gamma^0 a \frac{\partial}{\partial t} i \Delta[\xi,M^2_{\pm}](x,x) \; ,
\end{equation}
which follows from the mode expansion of the scalar propagator. Note that 
$i [\mbox{}_i S_i](x;x)$ is real even though each $i\Delta[\xi_c,M_{\pm}^2](x;x)$
is complex.

\subsection{Scalar Mode Amplitude}

It is most convenient to represent the scalar propagator as a spatial Fourier
mode sum,
\begin{eqnarray}
\lefteqn{i\Delta[\xi_c,M^2_{\pm}](x;x') = \int \!\! \frac{d^{D-1}k}{(2\pi)^{D-1}} 
\, e^{i \vec{k} \cdot \Delta \vec{x}} \Bigl\{ \theta(\Delta t) u(t,k,M_{\pm})
u^*(t',k,M_{\mp}) } \nonumber \\ 
& & \hspace{5.9cm} + \theta(-\Delta t) u^*(t,k,M_{\mp}) u(t',k,M_{\pm}) \Bigr\} , 
\qquad \label{scalar propagator}
\end{eqnarray}
where $\Delta \vec{x} \equiv \vec{x} - \vec{x}'$ and $\Delta t \equiv t - t'$.
Here $u(t,k,M_{\pm})$ is the plane wave mode function for a scalar of mass $M_{\pm}$ 
and conformal coupling $\xi_c \times R = \frac14 (\frac{D-2}{D-1}) \times (D-1) 
(D - 2 \epsilon) H^2$ obeying the equations,
\begin{eqnarray}
\Biggl[ \frac{d^2}{dt^2} + (D\!-\!1) H \frac{d}{dt} + \frac{k^2}{a^2} + M^2_{\pm}
+ \Bigl( \frac{D}2 \!-\! 1\Bigr) \!\Bigl( \frac{D}{2} \!-\! \epsilon\Bigr) \!\Biggr] 
u(t,k,M_{\pm}) &\!\!\! = \!\!\!& 0 \; , \qquad \label{ueqn} \\
u(t,k,M_{\pm}) \dot{u}^*(t,k,M_{\mp}) - \dot{u}(t,k,M_{\pm}) u^*(t,k,M_{\mp}) 
&\!\!\! = \!\!\!& \frac{i}{a^{D-1}} . \qquad \label{Wronskian}
\end{eqnarray}
Note that $u(t,k,M_{\pm})$ and $u^*(t,k,M_{\mp})$ obey the same equations,
which is why $u(t,k,M_{\pm})$ is paired with $u^*(t,k,M_{\mp})$ in the mode
sum (\ref{scalar propagator}) and the Wronskian (\ref{Wronskian}). Although exact 
solutions to (\ref{ueqn}) are not known for a general inflationary background
(\ref{geometry}), the Haddamard condition can be used to provide the initial 
conditions needed define a unique solution,
\begin{equation}
\frac{k}{a(t_i)} \gg Re(M_{\pm}) ,H(t_i) \qquad \Longrightarrow \qquad 
u(t,k,M_{\pm}) \longrightarrow \frac{\exp[-i \int_{t_i}^{t} 
\frac{k dt'}{a(t')}]}{\sqrt{2 k a^2(t)}} \; . \label{initial}
\end{equation}

Expression (\ref{fermiontrace}) only involves coincident scalar propagators,
\begin{equation}
i\Delta[\xi_c,M_{\pm}](x;x) = \int \!\! \frac{d^{D-1}k}{(2\pi)^{D-1}} \, 
u(t,k,M_{\pm}) u^*(t,k,M_{\mp}) \; .\label{scalar prop mode int}
\end{equation}
which are integrals of the complex product $u(t,k,M_{\pm}) u^*(t,k,M_{\mp})$. 
Because just this product is required we will infer an equation for it and 
then derive approximate solutions. It will also simplify the analysis if we 
change the evolution parameter from co-moving time $t$ to the dimensionless 
number of e-foldings from the beginning of inflation,
\begin{equation}
n \equiv \ln\Bigl[ \frac{a(t)}{a(t_i)}\Bigr] \quad \Longrightarrow \quad
\frac{d}{dt} = H \frac{d}{dn} \quad , \quad \frac{d^2}{dt^2} = H^2 \Bigl[
\frac{d^2}{dn^2} - \epsilon \frac{d}{dn}\Bigr] \; , 
\end{equation}
and extract factors of $\sqrt{8\pi G}$ to render the various parameters 
dimensionless,
\begin{equation}
\kappa \equiv \sqrt{8\pi G} \!\times\! k \;\; , \;\; 
\chi(n) \equiv \sqrt{8 \pi G} \!\times\! H(t) \;\; , \;\; 
\mu^2 \equiv 8 \pi G \!\times\! {\rm Re}\Bigl(M^2_{\pm}\Bigr) \; .
\label{dimless}
\end{equation}
The natural dependent variable is ,
\begin{equation}
\mathcal{M}_{\pm}(n,\kappa,\mu) \equiv \ln\Biggl[\frac{u(t,k,M_{\pm}) 
\!\times\! u^*(t,k,M_{\mp})}{\sqrt{8\pi G}} \Biggr] \; . \label{mathcalM}
\end{equation}
A now-familiar series of steps converts the mode equation (\ref{ueqn}) and
the Wronskian (\ref{Wronskian}) into a single complex equation for
$\mathcal{M}_{\pm}(n,\kappa,\mu)$ \cite{Romania:2012tb,Brooker:2015iya},
\begin{eqnarray}
\lefteqn{\mathcal{M}''_{\pm} + \frac{{\mathcal{M}'_{\pm}}^2}{2} + 
(D \!-\! 1 \!-\! \epsilon) \mathcal{M}'_{\pm} + 
\frac{2 \kappa^2 e^{-2 n}}{\chi^2} } \nonumber \\
& & \hspace{2.5cm} + \frac{2 \mu^2}{\chi^2} \mp \frac{2 i \mu}{\chi} + 
(D\!-\!2) \Bigl(\frac{D}{2} \!-\! \epsilon\Bigr) - \frac{ e^{-2 (D-1) n 
- 2 \mathcal{M}_{\pm}}}{2 \chi^2} = 0 \, , \qquad \label{Meqn}
\end{eqnarray}
where a prime denotes differentiation with respect to $n$. In the new 
variables the initial condition (\ref{initial}) is,
\begin{equation}
\mathcal{M}_{\pm}(0,\kappa,\mu) = \ln\Bigl[ \frac1{2\kappa} \Bigr] \qquad , 
\qquad \mathcal{M}_{\pm}'(0,\kappa,\mu) = -2 \; . \label{initial conditions M}
\end{equation}
 
\subsection{Two Phases}

Because equation (\ref{Meqn}) cannot be solved exactly for a general 
inflationary geometry (\ref{geometry}) we will solve $\mathcal{M}_{\pm}(n,\kappa,\mu)$
numerically to motivate, and then to validate, analytic approximations. Numerical 
solution obviously requires explicit formulae for the dimensionless geometrical 
parameters $\chi(n)$ and $\epsilon(n)$. We construct these using the natural
dimensionless expression of the scalar evolution equation (\ref{scalareqn}),
\begin{equation}
\phi'' + (3 \!-\! \epsilon) \phi' + \frac{U'(\phi)}{\chi^2} = 0 \; , 
\label{Infl4}
\end{equation}
where $\phi \equiv \sqrt{8 \pi G} \times \varphi$ and $U(\phi) \equiv
(8\pi G)^2 \times V(\varphi)$. The dimensionless expressions of the geometrical
relations (\ref{E1}-\ref{E2}) are
\begin{equation}
\chi^2 = \frac{U}{3 \!-\! \frac12 {\phi'}^2} \qquad , \qquad \epsilon = \frac12
{\phi'}^2 \; . \label{Infl5}
\end{equation}
For simplicity we have chosen the simple quadratic model (\ref{Vclass}), which 
corresponds to $U(\phi) = \frac12 c^2 \phi^2$. For this model the slow roll 
approximation gives,
\begin{equation}
\phi(n) \simeq \sqrt{\phi_0^2 \!-\! 4n} \quad , \quad \chi(n) \simeq \frac{c}{\sqrt{6}} 
\sqrt{\phi_0^2 \!-\! 4n} \quad , \quad \epsilon(n) \simeq \frac{2}{\phi_0^2 \!-\! 4 n}
\; , \label{slowroll}
\end{equation}
where we have abused the notation slightly by regarding the first slow roll 
parameter as a function of $n$ rather than $t$. Note also the relation $\chi(n) 
\simeq \chi_0 \sqrt{1 - 4n/\phi_0^2}$. By choosing the initial value $\phi_0 = 15$ 
we get about 50 e-foldings of inflation. With $c = 7.1 \times 10^{-6}$ this model 
agrees with the observed scalar amplitude and spectral index, however, its prediction
for the tensor-to-scalar ratio is far too big \cite{Aghanim:2018eyx}. At the end of
the section we will demonstrate that our results pertain as well for viable models.

\begin{figure}[H]
	\centering
	\begin{subfigure}[b]{0.33\textwidth}
		\centering
		\includegraphics[width=\textwidth]{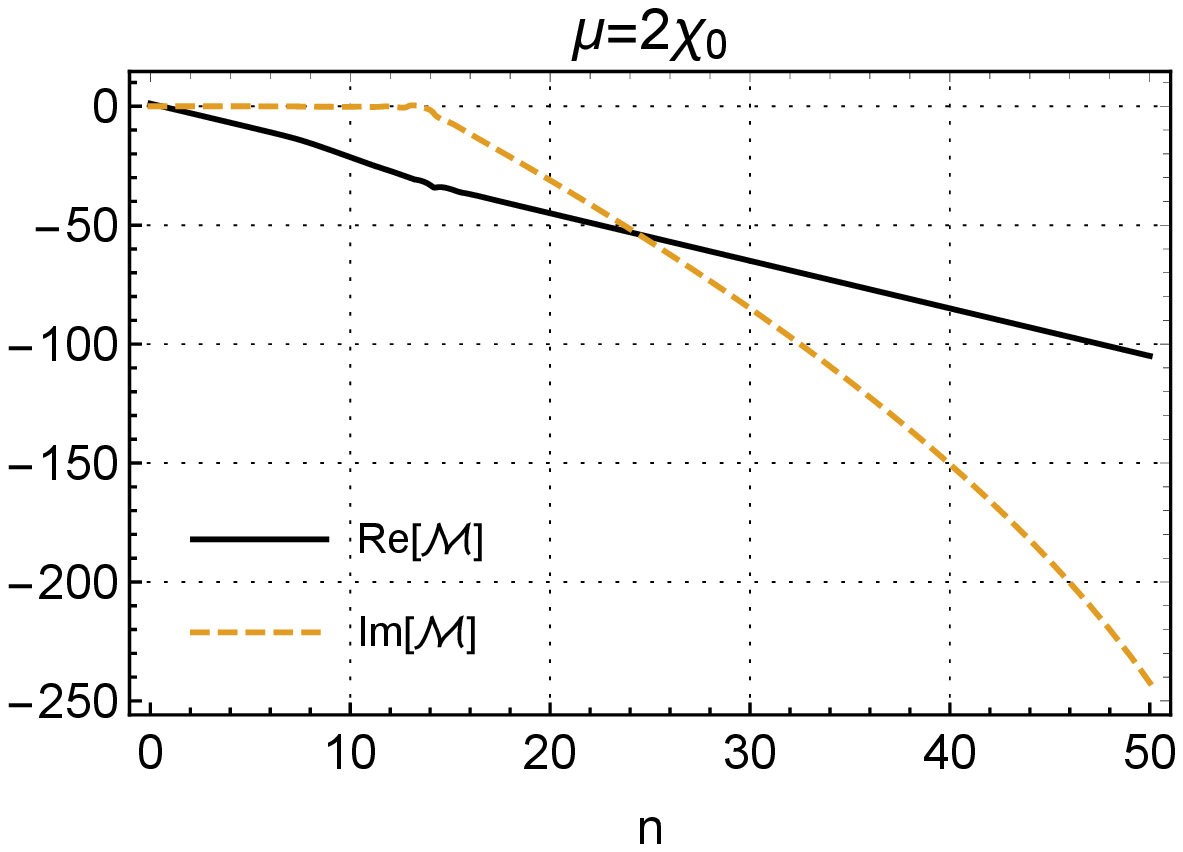}
		\caption{$\mu = 2 \chi_0$}
	\end{subfigure}
	\begin{subfigure}[b]{0.33\textwidth}
		\centering
		\includegraphics[width=\textwidth]{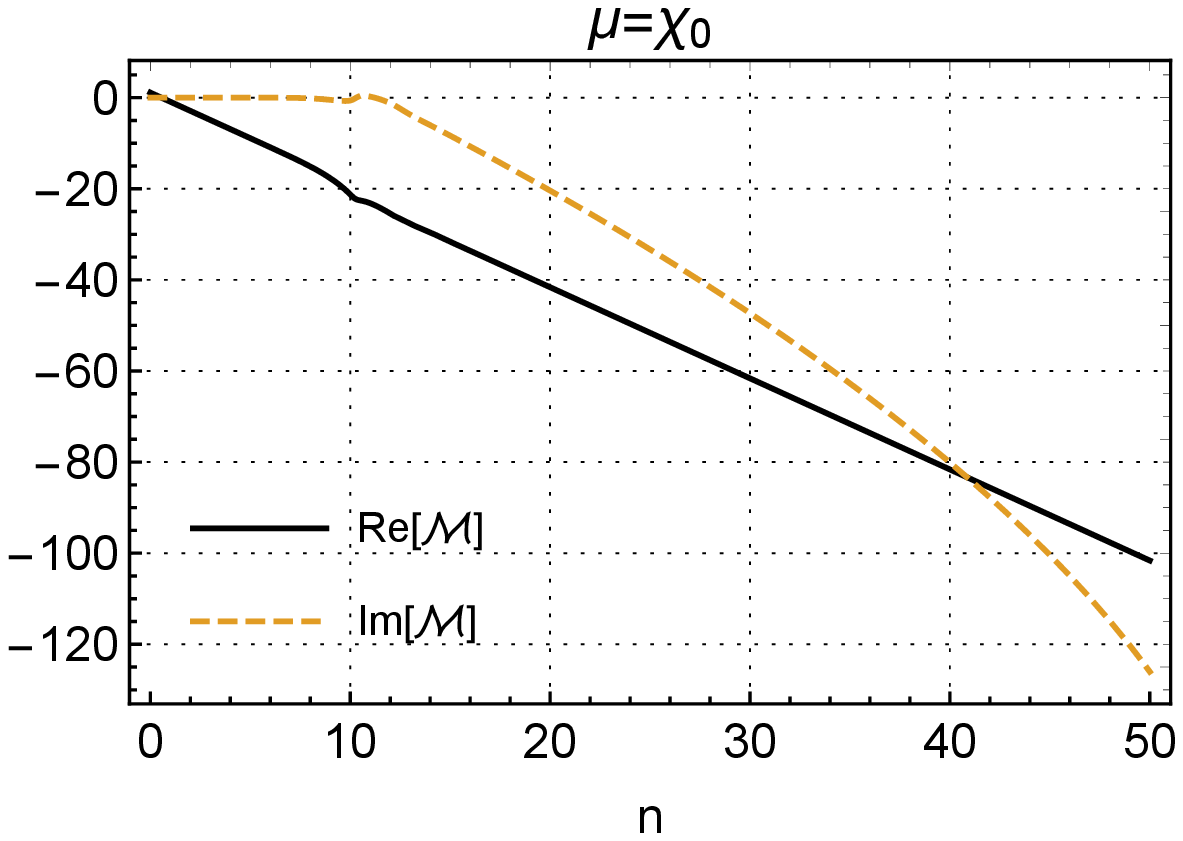}
		\caption{$\mu = \chi_0$}
	\end{subfigure}
	\begin{subfigure}[b]{0.33\textwidth}
		\centering
		\includegraphics[width=\textwidth]{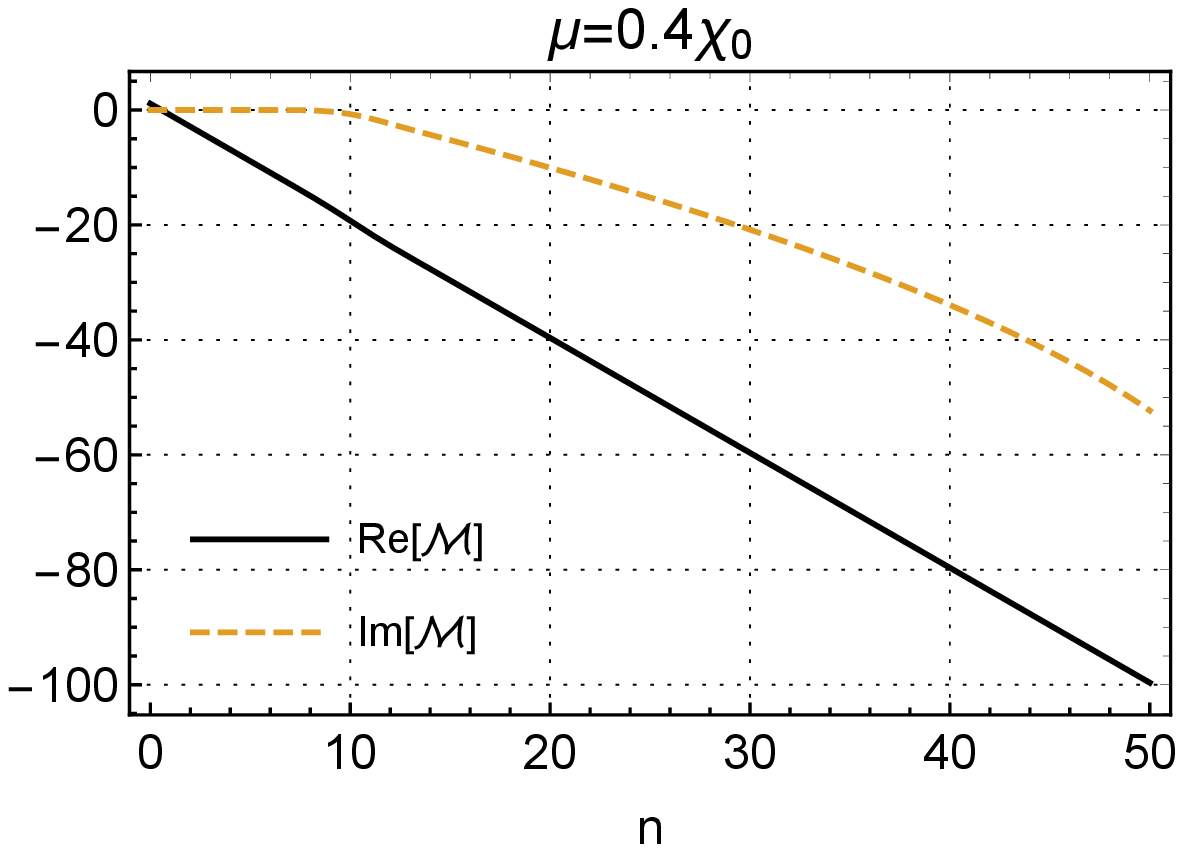}
		\caption{$\mu = \frac25 \chi_0$}
	\end{subfigure}%
	\caption{\footnotesize Plots of the complex amplitude 
	$\mathcal{M}_{-}(n,\kappa,\mu)$ versus the e-folding $n$ for 
	$\kappa = 3800 \chi_0$ (which experiences horizon crossing at 
	$n_{\kappa} \simeq 8.3$) at $\mu = 2 \chi_0$ (left), $\mu = \chi_0$ 
	(middle) and $\mu = \frac25 \chi_0$ (right). In each case real part 
	is in solid black while the imaginary part is in long dashed yellow.}
	\label{MuBIG-figure}
\end{figure}

\begin{figure}[H]
	\centering
	\begin{subfigure}[b]{0.33\textwidth}
		\centering
		\includegraphics[width=\textwidth]{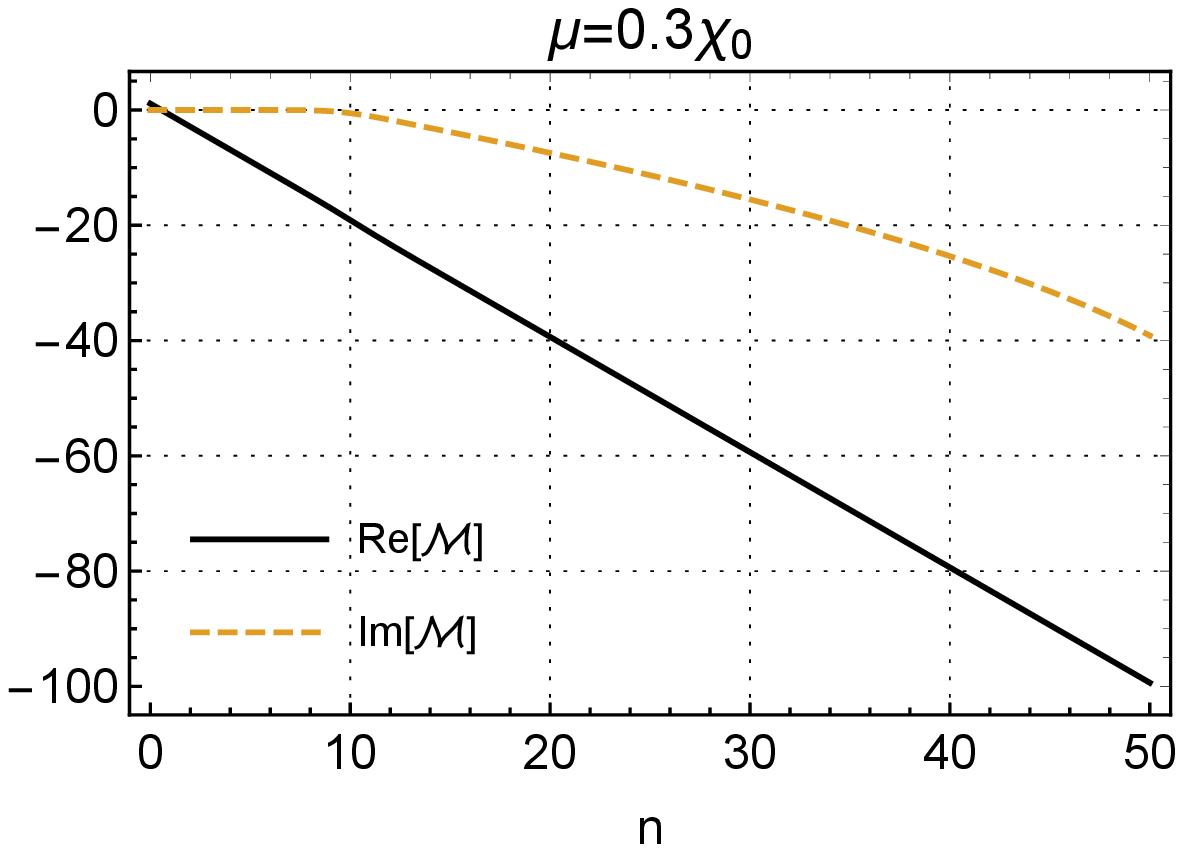}
		\caption{$\mu = \frac3{10} \chi_0$}
	\end{subfigure}
	\begin{subfigure}[b]{0.33\textwidth}
		\centering
		\includegraphics[width=\textwidth]{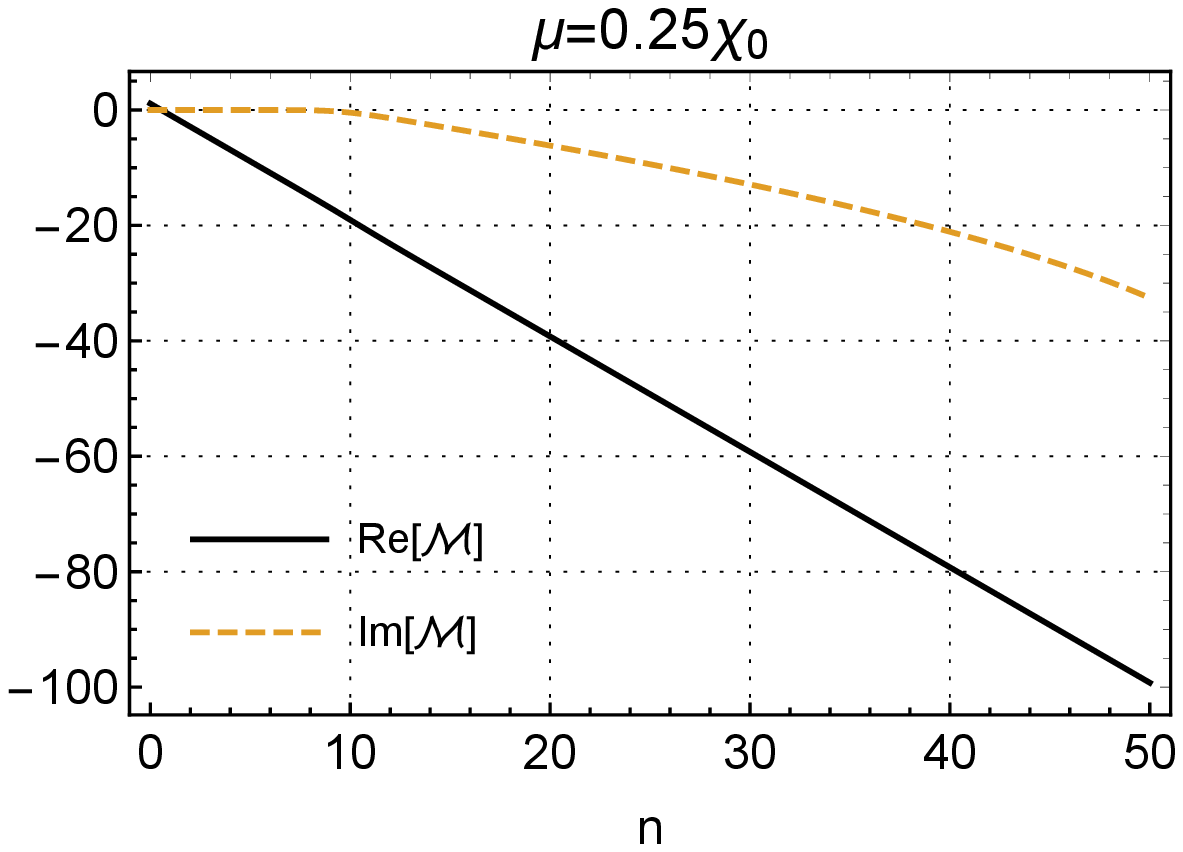}
		\caption{$\mu = \frac14 \chi_0$}
	\end{subfigure}
	\begin{subfigure}[b]{0.33\textwidth}
		\centering
		\includegraphics[width=\textwidth]{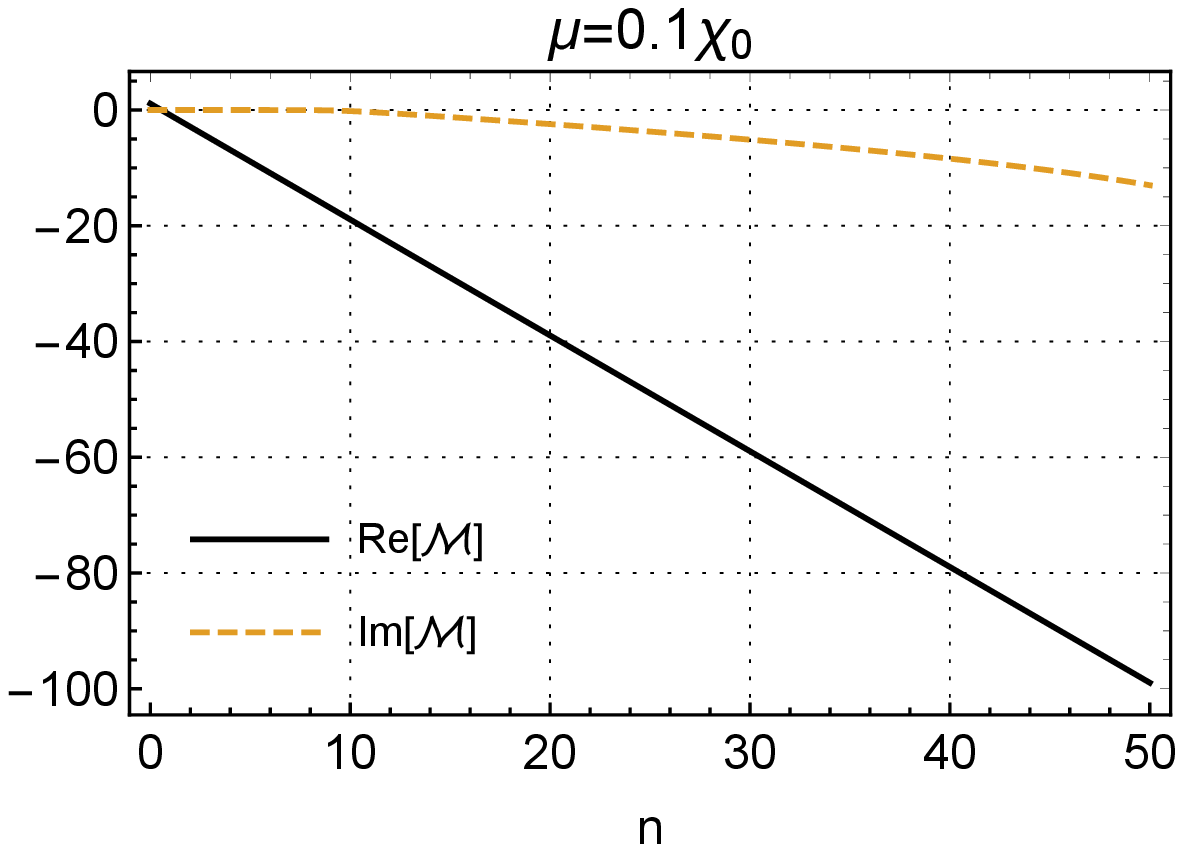}
		\caption{$\mu = \frac1{10} \chi_0$}
	\end{subfigure}%
	\caption{\footnotesize Plots of the complex amplitude 
	$\mathcal{M}_{-}(n,\kappa,\mu)$ versus the e-folding $n$ for 
	$\kappa = 3800 \chi_0$ (which experiences horizon crossing at $n_{\kappa} 
	\simeq 8.3$) at $\mu = \frac3{10} \chi_0$ (left), $\mu = \frac14 \chi_0$ 
	(middle) and $\mu = \frac1{10} \chi_0$ (right). In each case real part 
	is in solid black while the imaginary part is in long dashed yellow.}
	\label{MuSMALL-figure}
\end{figure}

Figures \ref{MuBIG-figure} and \ref{MuSMALL-figure} give the evolution of 
$\mathcal{M}_{-}(n,\kappa,\mu)$ in the geometry (\ref{slowroll}) for $\kappa =
3800 \chi_0$ and six different values of $\mu$.
\noindent Because $\mathcal{M}_{-}(n,\kappa,\mu) = \mathcal{M}_{+}^*(n,\kappa,\mu)$ 
is complex, results are plotted for both the real part and imaginary parts.
\noindent Because the initial conditions (\ref{initial conditions M}) are purely 
real, the imaginary part is zero for small $n$, and then builds up after 
horizon crossing $\kappa = e^{n_{\kappa}} \chi(n_{\kappa})$. The imaginary part 
is larger, and begins growing earlier, for larger $\mu$ because it is driven by 
the $\mp 2 i \mu/\chi(n)$ term in equation (\ref{Meqn}). In contrast, the real 
part is almost the same for all values of $\mu$.

\begin{figure}[H]
	\centering
	\begin{subfigure}[b]{0.33\textwidth}
		\centering
		\includegraphics[width=\textwidth]{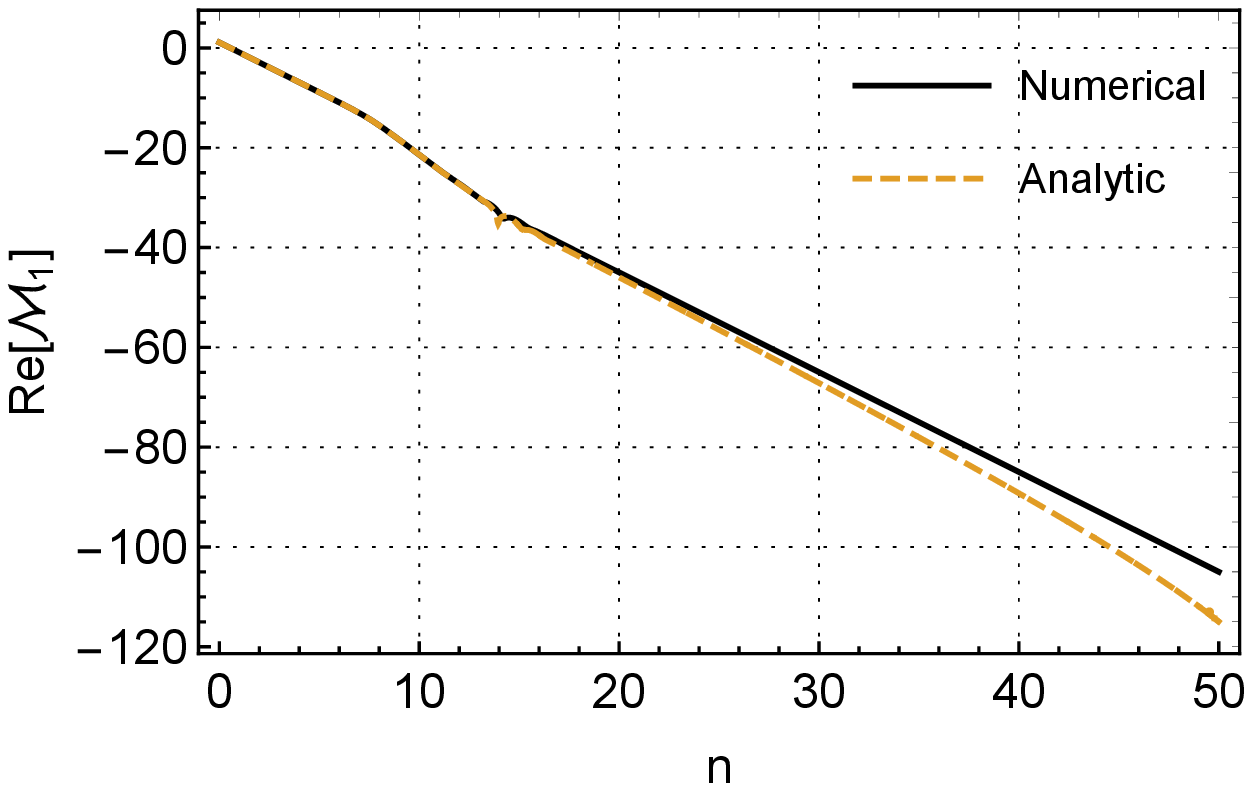}
		\caption{$\mu = 2 \chi_0$}
	\end{subfigure}
	\begin{subfigure}[b]{0.33\textwidth}
		\centering
		\includegraphics[width=\textwidth]{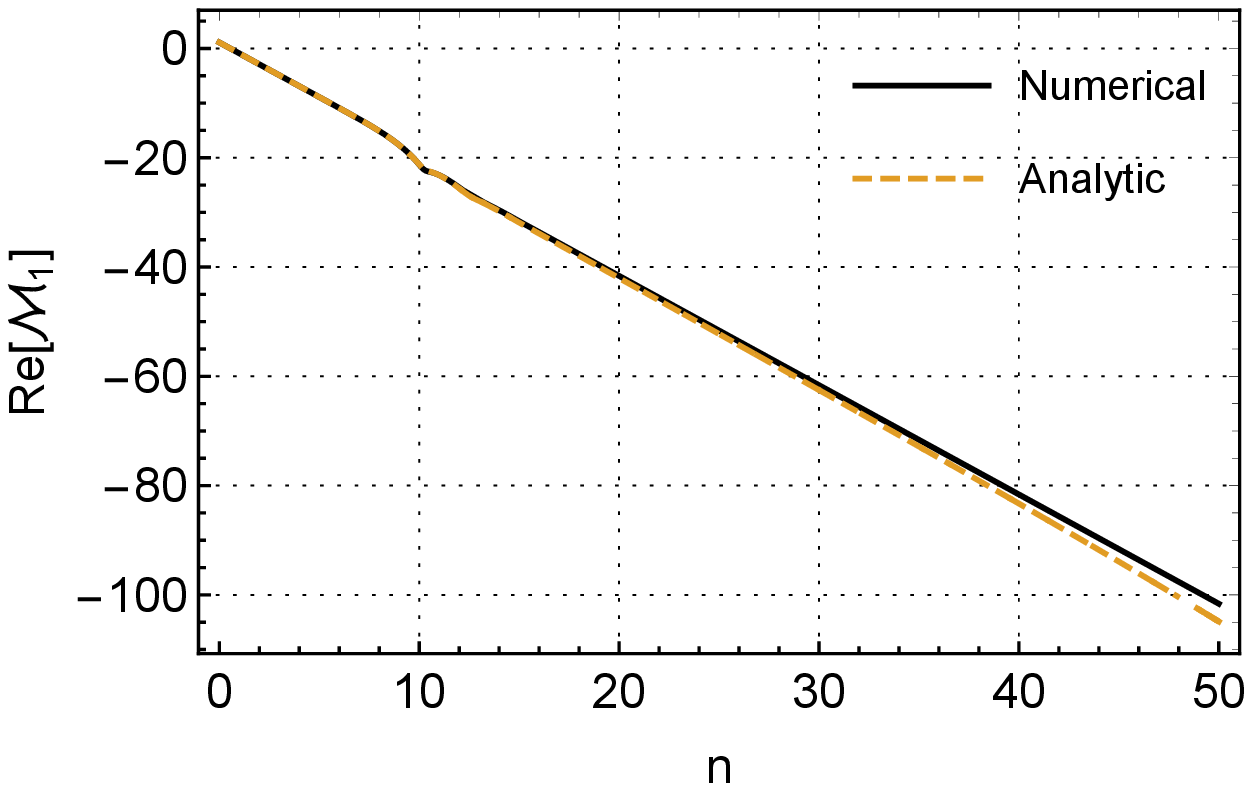}
		\caption{$\mu = \chi_0$}
	\end{subfigure}
	\begin{subfigure}[b]{0.33\textwidth}
		\centering
		\includegraphics[width=\textwidth]{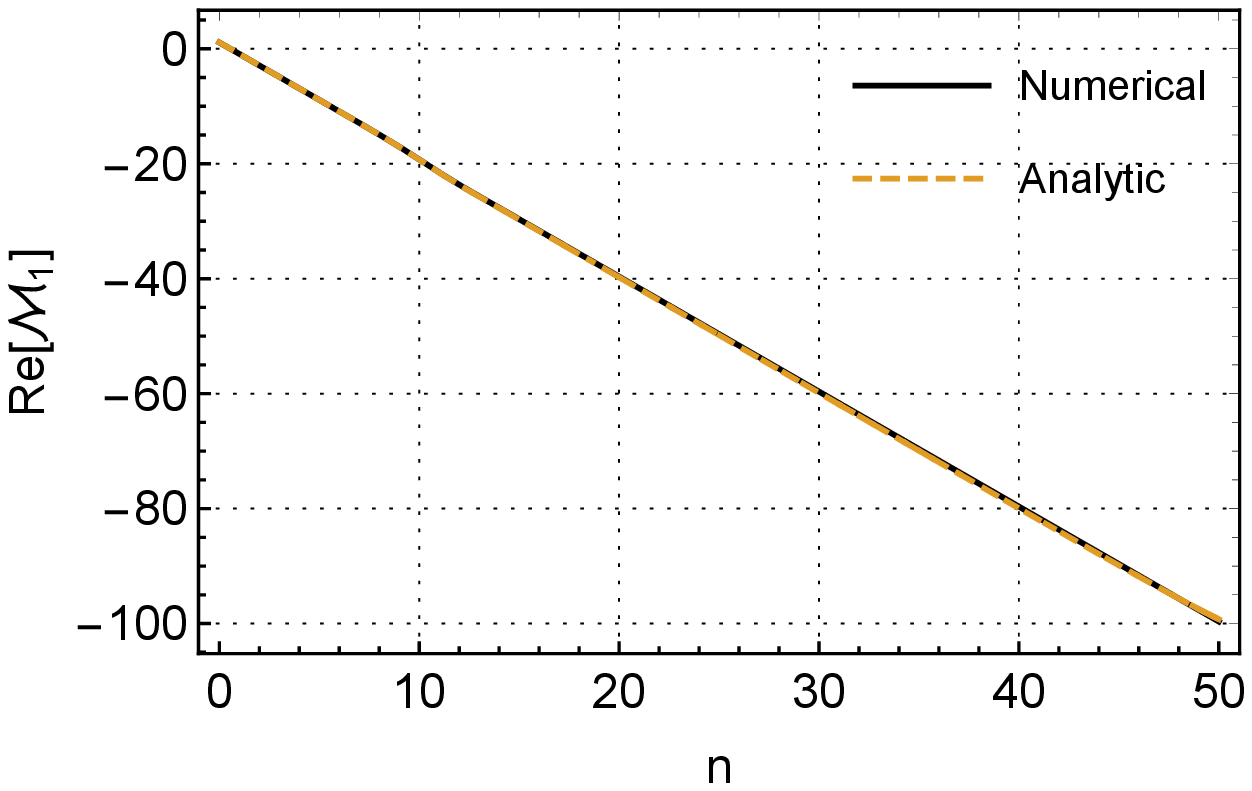}
		\caption{$\mu = \frac25 \chi_0$}
	\end{subfigure}%
	\caption{\footnotesize Comparing the real parts of $\mathcal{M}_{-}(n,3800 \chi_0,\mu)$ 
	and $\mathcal{M}_{-1}(n,3800 \chi_0,\mu)$ for $\mu = 2 \chi_0$ (left), $\mu = 
	\chi_0$ (middle) and $\mu = \frac25 \chi_0$ (right). In each case the numerical
	solution is solid black while the approximation is long dashed yellow.}
	\label{M1real-figure}
\end{figure}

\begin{figure}[H]
	\centering
	\begin{subfigure}[b]{0.33\textwidth}
		\centering
		\includegraphics[width=\textwidth]{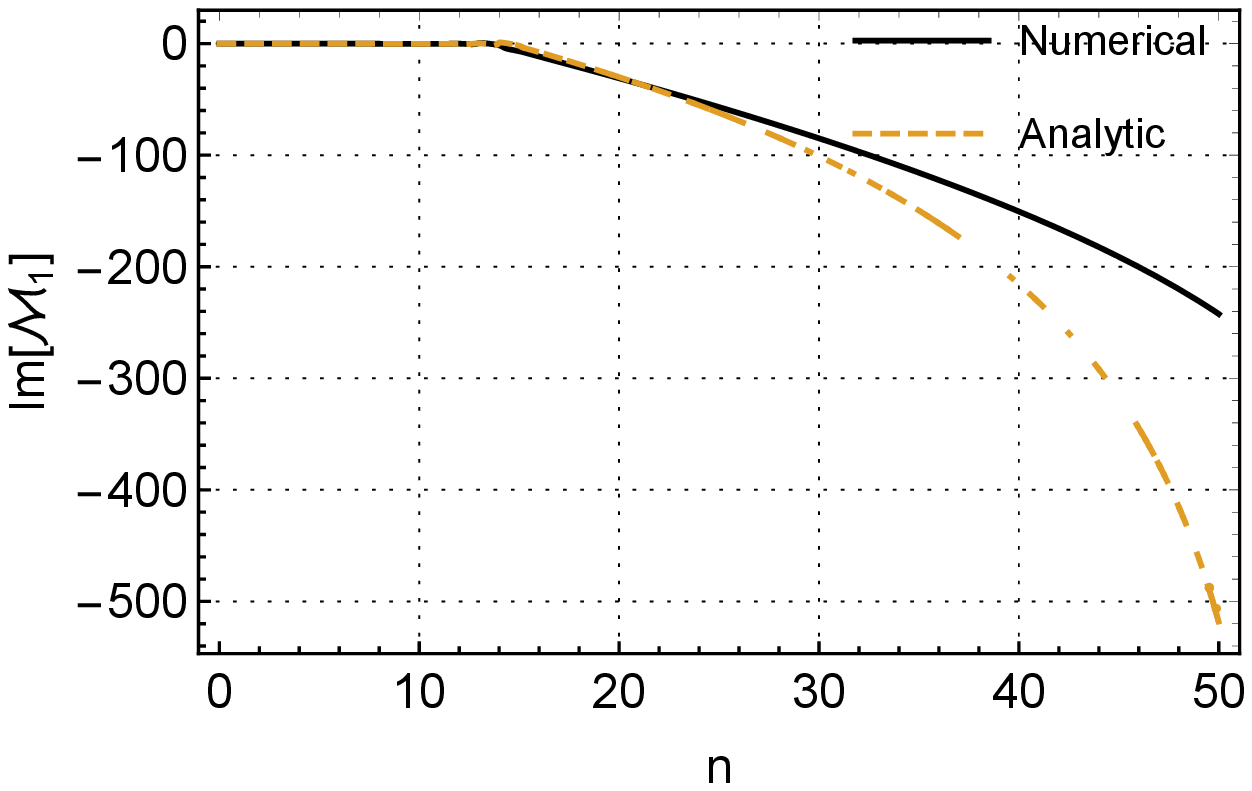}
		\caption{$\mu = 2 \chi_0$}
	\end{subfigure}
	\begin{subfigure}[b]{0.33\textwidth}
		\centering
		\includegraphics[width=\textwidth]{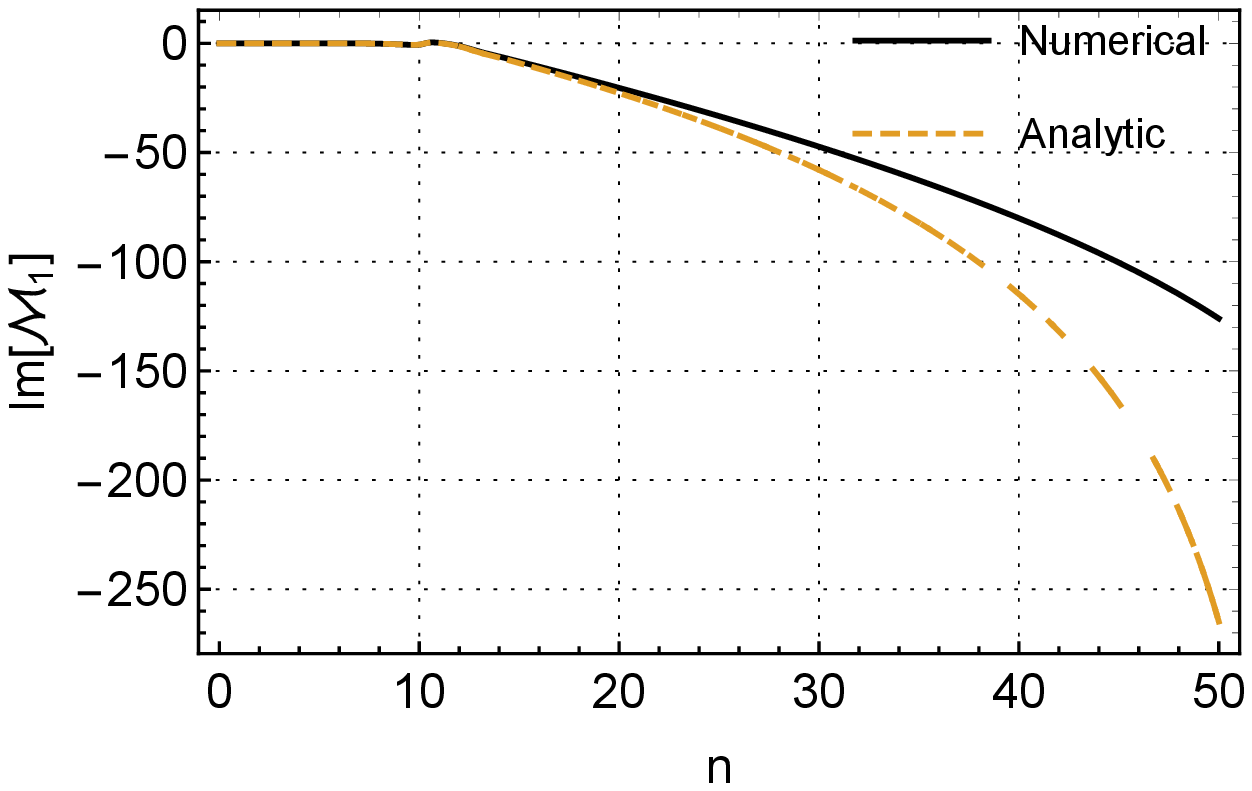}
		\caption{$\mu = \chi_0$}
	\end{subfigure}
	\begin{subfigure}[b]{0.33\textwidth}
		\centering
		\includegraphics[width=\textwidth]{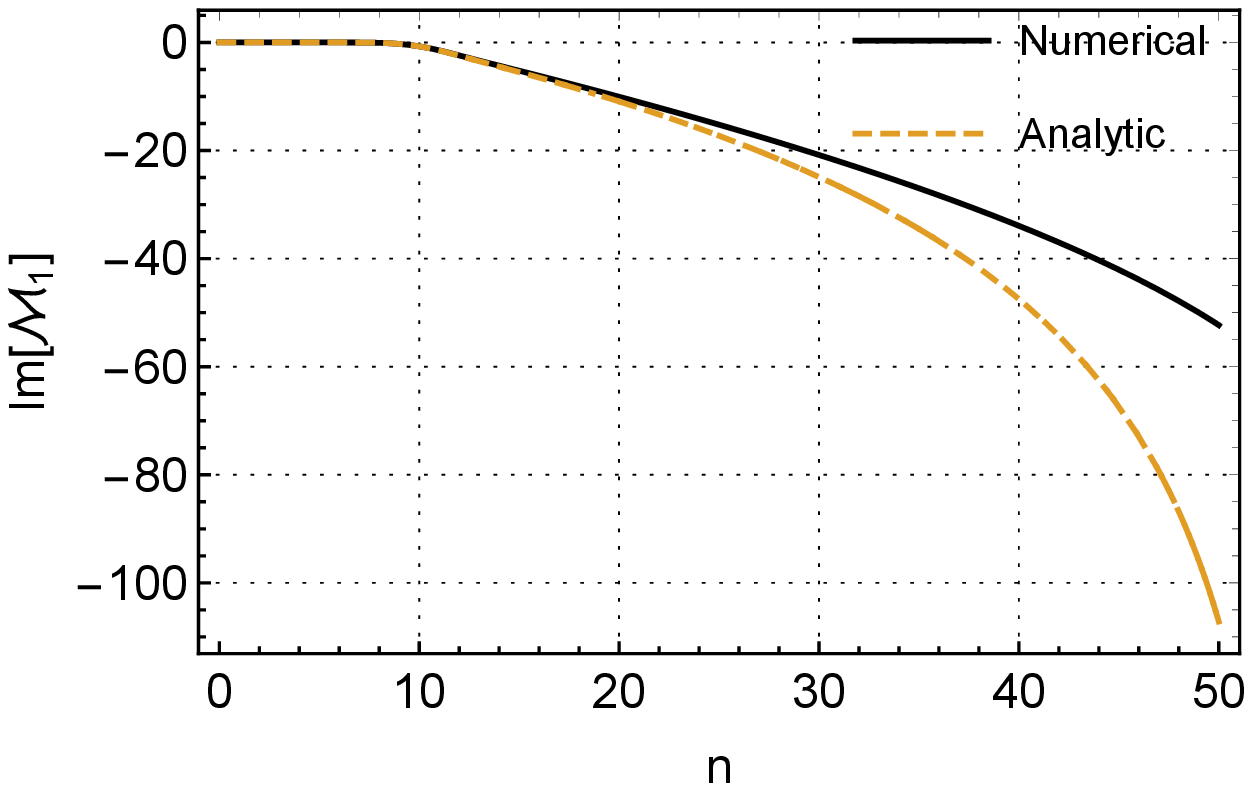}
		\caption{$\mu = \frac25 \chi_0$}
	\end{subfigure}%
	\caption{\footnotesize Comparing the imaginary parts of $\mathcal{M}_{-}(n,3800 
	\chi_0,\mu)$ and $\mathcal{M}_{-1}(n,3800 \chi_0,\mu)$ for $\mu = 2 \chi_0$ (left), 
	$\mu = \chi_0$ (middle) and $\mu = \frac25 \chi_0$ (right). In each case the 
	numerical solution is solid black while the approximation is long dashed yellow.}
	\label{M1imaginary-figure}
\end{figure}

Up until horizon crossing, and even somewhat later, it is generally valid to 
use the Hankel function solution that pertains in the far ultraviolet,\footnote{
This approximation also becomes exact for the case of constant $\epsilon$ with 
$\mu \propto \chi$ \cite{Janssen:2009pb}.}
\begin{equation}
\mathcal{M}_1(n,\kappa,\mu) \equiv \ln\ls \frac{\frac{\pi}{2} z(n,\kappa)}{2 
\kappa e^{(D-2) n}} \!\times\! H^{(1)}_{\nu_{\pm}(n,\mu)}\Bigl(z(n,\kappa) \Bigr) 
\!\times\! \Bigr[H^{(1)}_{\nu_{\mp}(n,\mu)}\Bigl(z(n,\kappa) \Bigr) \Bigr]^* \rs 
\; , \label{M1def}
\end{equation}
where we define,
\begin{equation}
\nu_{\pm}(n,\mu) \equiv \sqrt{\frac14 - \frac{(\mu^2 \!\mp\! i \chi \mu)}{(1 \!-\! 
\epsilon)^2 \chi^2}} \qquad , \qquad z(n,\kappa) \equiv \frac{\kappa e^{-n}}{(1 \!-\! 
\epsilon) \chi} \; . \label{hankel index}
\end{equation}
Figures~\ref{M1real-figure} and \ref{M1imaginary-figure} compare the real and 
imaginary parts of $\mathcal{M}_{-1}(n,\kappa,\mu)$ with the numerical evolution of 
$\mathcal{M}_{-}(n,\kappa,\mu)$. 
\noindent Note that the real part of the $\mathcal{M}_{-1}$ approximation is close to 
the numerical solution even long after horizon crossing, and only differs visibly at
the near the end of inflation and for the largest values of $\mu$. In contrast, the
disagreement of the imaginary parts becomes visible between 20 and 30 e-foldings.

To see analytically that (\ref{M1def}) captures the far ultraviolet regime of 
$\kappa \gg e^n \chi(n)$, first write the exact result as the approximation plus 
a deviation, $\mathcal{M}_{\pm} \equiv \mathcal{M}_{\pm 1} + g_{\pm}$. Now extract 
the $z$-dependent part of the approximation as $\mathcal{M}_{\pm 1} \equiv -\ln(2\kappa) 
- (D - 2) n + \Delta \mathcal{M}_{\pm 1}$. The ultraviolet corresponds to large $z$ so 
we employ the large $z$ expansion of the Hankel function in $\Delta \mathcal{M}_{\pm 1}$,
\begin{equation}
e^{\Delta \mathcal{M}_{\pm 1}} = \frac{\pi z}{2} H^{(1)}_{\nu_{\pm}}(z) 
\Bigl(H^{(1)}_{\nu_{\mp}}(z) \Bigr)^* = 1 + \frac{ (\nu_{\pm}^2 \!-\! \frac14)}{2 z^2} 
+ \frac{3 (\nu_{\pm}^2 \!-\! \frac14) (\nu_{\pm}^2 \!-\! \frac94)}{8 z^4} + O\Bigl( 
\frac1{z^6}\Bigr) \; . \label{largez}
\end{equation}
Substituting the various expansions into equation (\ref{Meqn}) results in an
series for the deviation $g_{\pm}(n,\kappa,\mu)$ in powers of $(\chi e^n/\kappa)^2$,
\begin{equation}
g_{\pm} = \Bigl[ \frac{\epsilon (5 \!-\! 3 \epsilon) \mu (\mu \mp i \chi)}{4 \chi^2} 
\Bigr] \Bigl( \frac{\chi e^n}{\kappa}\Bigr)^4 + O\Biggl( 
\Bigl(\frac{\chi e^n}{\kappa}\Bigr)^6 \Biggr) . \label{gexp}
\end{equation}
Note that relation (\ref{gexp}) correctly predicts the trend we saw in Figure
\ref{M1imaginary-figure} that ${\rm Im}[\mathcal{M}_{-} - \mathcal{M}_{-1}]$ is 
positive. The fact that $\mathcal{M}_{\pm}(n,\kappa,\mu) = \mathcal{M}_{\pm 1}(n,
\kappa,\mu) + O(\kappa^{-4})$ is crucial in computing the coincident propagator 
because it means the $\mathcal{M}_{\pm 1}$ approximation includes all ultraviolet 
divergences,
\begin{eqnarray}
\lefteqn{ \int \!\! d\kappa \, \kappa^{D-2} e^{\mathcal{M_{\pm}}(n,\kappa,\mu)} = 
\frac{\chi^{D-2}}2 \int \!\! \frac{d\kappa}{\kappa} \Bigl( \frac{\kappa e^{-n}}{
\chi(n)}\Bigr)^{D-2} } \nonumber \\
& & \hspace{3cm} \times \Biggl\{ 1 - \Bigl[\frac{\mu (\mu \mp i \chi)}{2 \chi^2}
\Bigr] \Bigl( \frac{\chi(n)}{\kappa e^{-n}}\Bigr)^2 + O\Biggl( 
\Bigl(\frac{\chi(n)}{\kappa e^{-n}}\Bigr)^{4} \Biggr) \Biggr\} . \qquad \label{UVexp}
\end{eqnarray}
Hence we can dispense with dimensional regularization when approximating 
$\mathcal{M}_{\pm}(n,\kappa,\mu)$ for $n > n_{\kappa}$.

\begin{figure}[H]
	\centering
	\begin{subfigure}[b]{0.33\textwidth}
		\centering
		\includegraphics[width=\textwidth]{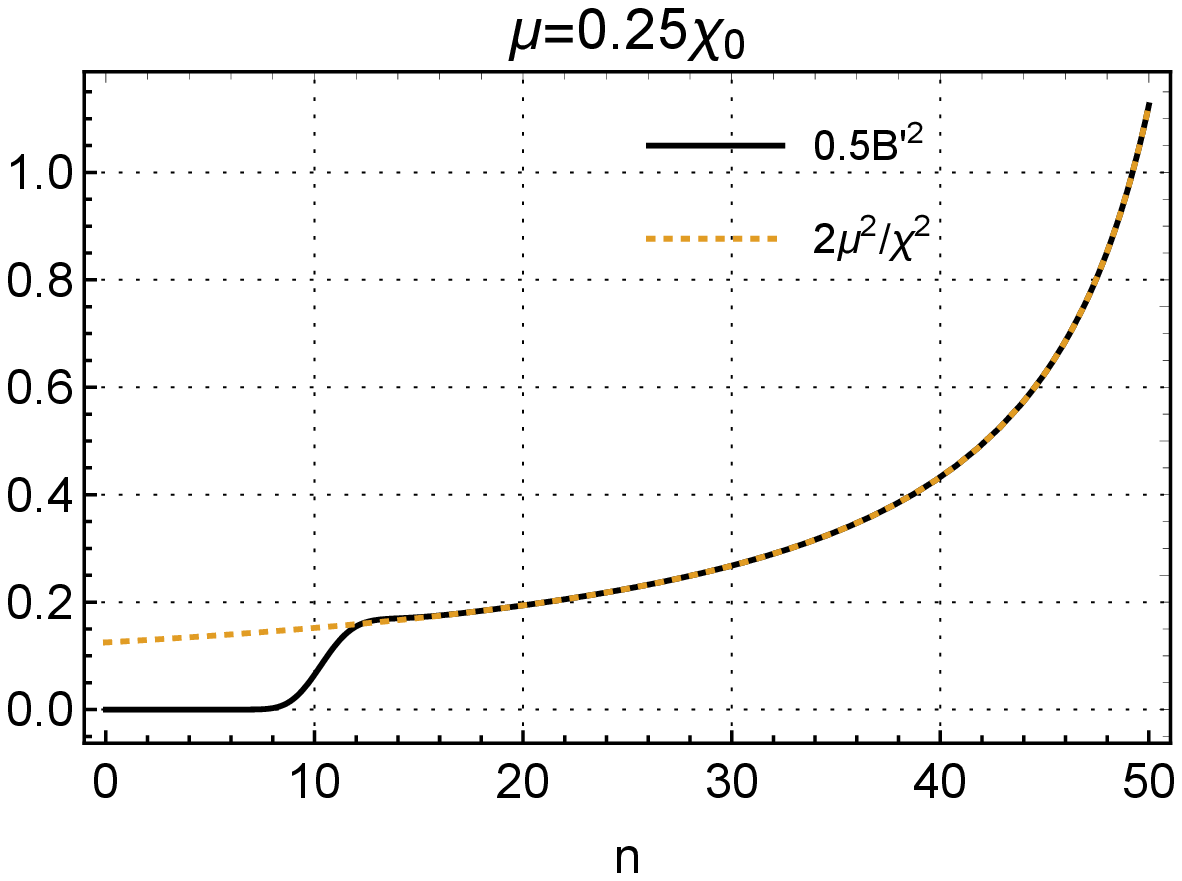}
		\caption{$\mu = \frac14 \chi_0$}
	\end{subfigure}
	\begin{subfigure}[b]{0.33\textwidth}
		\centering
		\includegraphics[width=\textwidth]{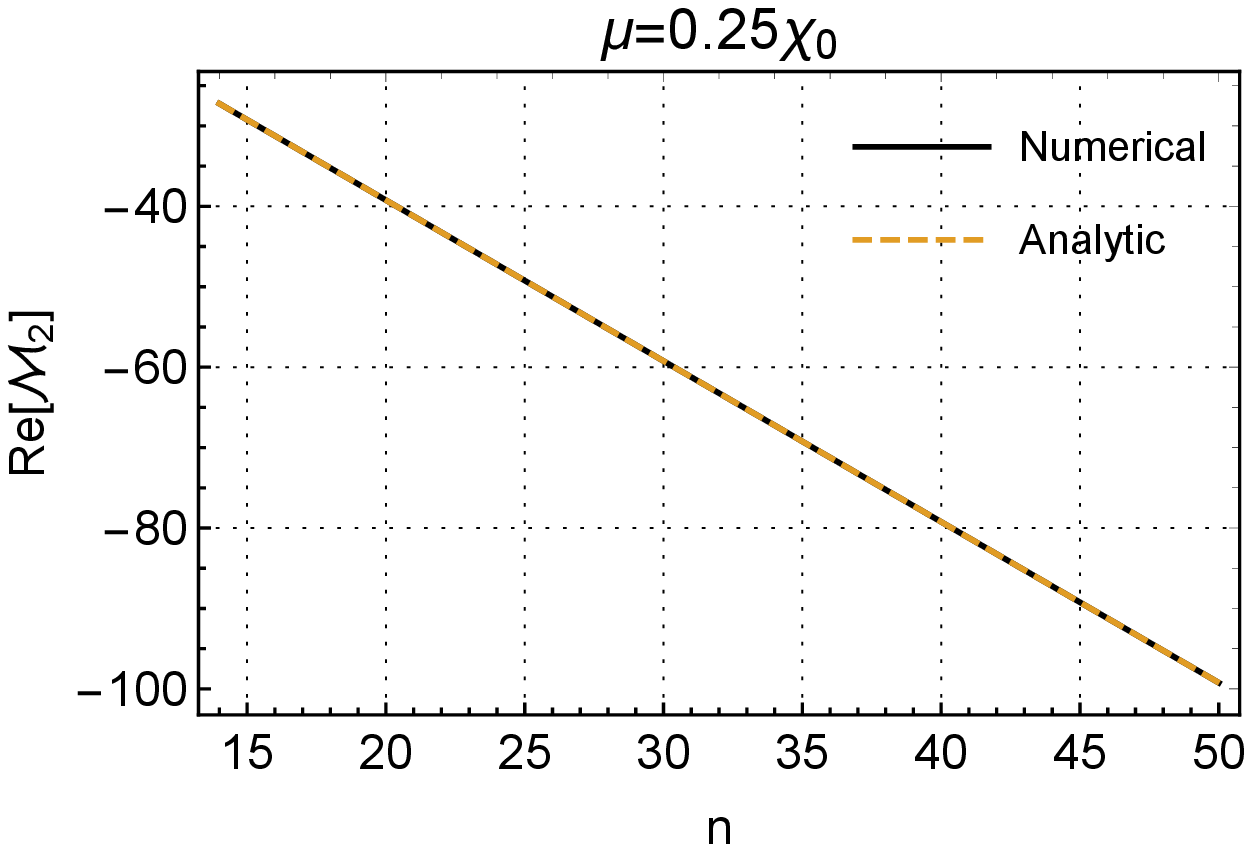}
		\caption{$\mu = \frac14 \chi_0$}
	\end{subfigure}
	\begin{subfigure}[b]{0.33\textwidth}
		\centering
		\includegraphics[width=\textwidth]{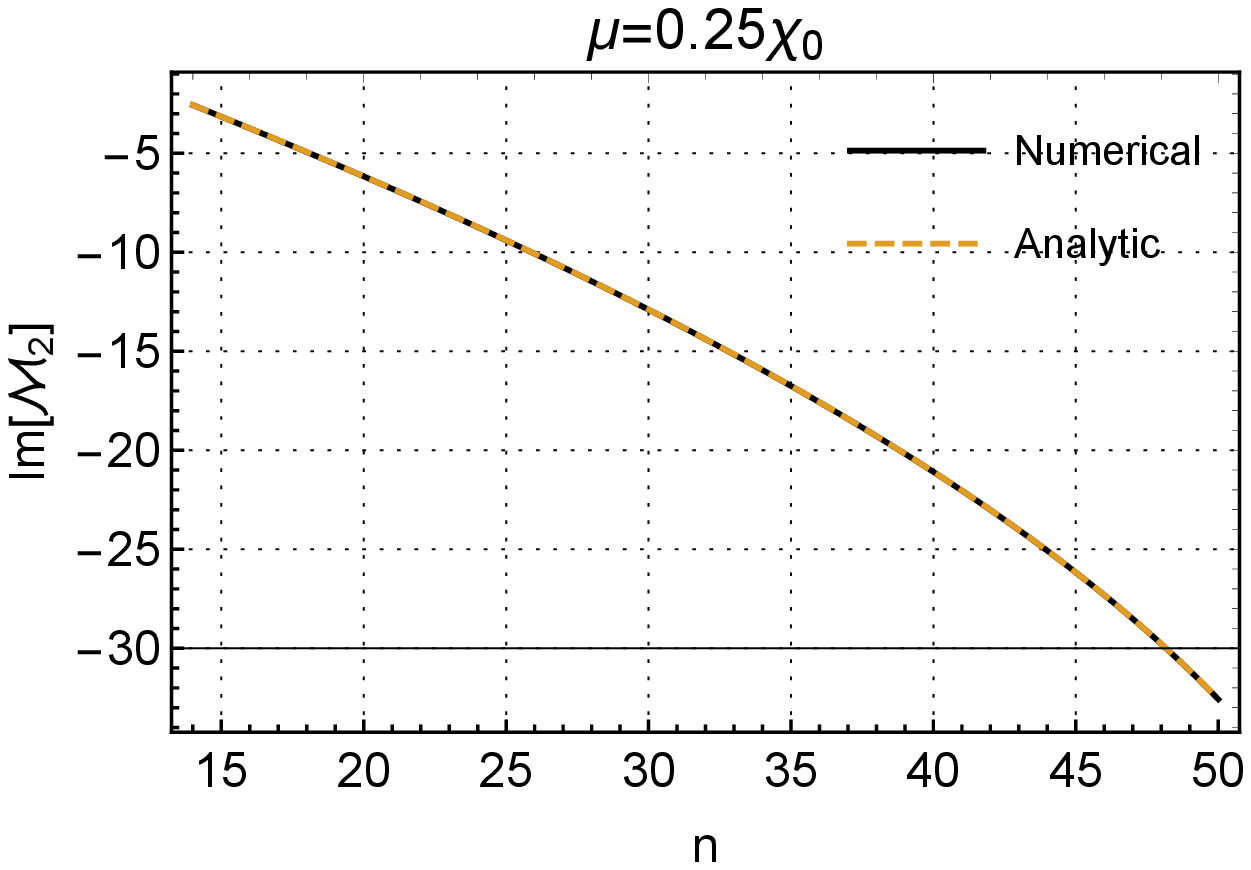}
		\caption{$\mu = \frac14 \chi_0$}
	\end{subfigure}%
	\caption{\footnotesize These plots justify the $\mathcal{M}_{-2}(n,\kappa,\mu)$
	approximation (\ref{M2def}) for the post-horizon amplitude. The left hand graph
	compares $\frac12 {B_{-}'}^2$ with $2 \mu^2/\chi^2$ for $\kappa = 3800 \chi_0$ 
	and $\mu = \frac14 \chi_0$. The middle and right hand graphs compare the real
	and imaginary parts of $\mathcal{M}_{-}(n,3800 \chi_0,\frac14 \chi_0)$ with the
	approximation (\ref{M2def}).}
	\label{M2approx-figure}
\end{figure}

Equation (\ref{Meqn}) contains seven terms, of which the 4th ($2 
(\kappa e^{-n}/\chi)^2$) and the 7th ($-\exp[-6n - 2\mathcal{M}_{\pm}]/2 \chi^2$) 
dominate at the beginning of inflation. During this initial phase 
$\mathcal{M}_{\pm}(n,\kappa,\mu)$ falls off like $-2n$. After horizon crossing the 
4th and 7th terms rapidly redshift to zero and the equation becomes 
approximately,
\begin{equation}
\mathcal{M}_{\pm}'' + \frac12 {\mathcal{M}_{\pm}'}^2 + (3 \!-\! \epsilon) 
\mathcal{M}_{\pm}' + \frac{2 \mu^2}{\chi^2} \mp \frac{2 i \mu}{\chi} +
4 \!-\! 2\epsilon \simeq 0 \; . \label{Meqn approx}
\end{equation}
It is illuminating to break relation (\ref{Meqn approx}) up into real and imaginary 
parts with the substitution $\mathcal{M}_{\pm} \equiv A_{\pm} + i B_{\pm}$,
\begin{eqnarray}
A_{\pm}'' + \frac12 \Bigl( {A_{\pm}'}^2 \!-\! {B_{\pm}'}^2\Bigr) + (3 \!-\! 
\epsilon) A_{\pm}' +\frac{2 \mu^2}{\chi^2} + 4 \!-\! 2 \epsilon & \simeq & 0 \; , 
\label{Aeqn} \\
B_{\pm}'' + A_{\pm}' B_{\pm}' + (3 \!-\! \epsilon) B_{\pm}' \mp 
\frac{2 \mu}{\chi} & \simeq & 0 \; . \label{Beqn}
\end{eqnarray}
If we neglect $\epsilon$ and the second derivatives, relations
(\ref{Aeqn}-\ref{Beqn}) become,
\begin{eqnarray}
\frac12 \Bigl( A_{\pm}' \!+\! 2\Bigr) \Bigl( A_{\pm}' \!+\! 4\Bigr) & \approx &
\frac12 \Bigl[ {B_{\pm}'}^2 - \Bigl(\frac{2 \mu}{\chi}\Bigr)^2 \Bigr] \; ,
\label{Aeqn2} \\
\Bigl(3 \!+\! A_{\pm}'\Bigr) B_{\pm}' & \approx & \pm \frac{2 \mu}{\chi} \; .
\label{Beqn2}
\end{eqnarray}
The right hand side of (\ref{Aeqn2}) can be re-written in a suggestive form,
\begin{equation}
\frac12 \Bigl[ {B_{\pm}'}^2 - \Bigl(\pm \frac{2 \mu}{\chi}\Bigr)^2 \Bigr] =
\frac{[ (3 \!+\! A_{\pm}')^2 {B_{\pm}'}^2 - (\frac{2\mu}{\chi})^2]}{2
(3 \!+\! A_{\pm}')^2} + \frac{ (A_{\pm}' \!+\! 2) (A_{\pm}' \!+\! 4)}{2
(3 \!+\! A_{\pm}')^2} \Bigl(\frac{2\mu}{\chi}\Bigr)^2 . \label{RHS}
\end{equation}
Substituting (\ref{RHS}) in (\ref{Aeqn2}) reveals two solutions to the system
(\ref{Aeqn2}-\ref{Beqn2}),
\begin{eqnarray}
A_{\pm}' \approx -2 \qquad & \Longrightarrow & \qquad B_{\pm}' \approx \pm 
\frac{2\mu}{\chi} \; , \label{rightsol} \\
A_{\pm}' \approx -4 \qquad & \Longrightarrow & \qquad B_{\pm}' \approx \mp 
\frac{2 \mu}{\chi} \; . \label{wrongsol}
\end{eqnarray}
The second solution (\ref{wrongsol}) is ruled out by virtue of not being 
consistent with the neglect of the final term in (\ref{Meqn}). The left hand
graph of Figure~\ref{M2approx-figure} establishes that (\ref{rightsol}) is the
correct solution.

\begin{figure}[H]
	\centering
	\begin{subfigure}[b]{0.33\textwidth}
		\centering
		\includegraphics[width=\textwidth]{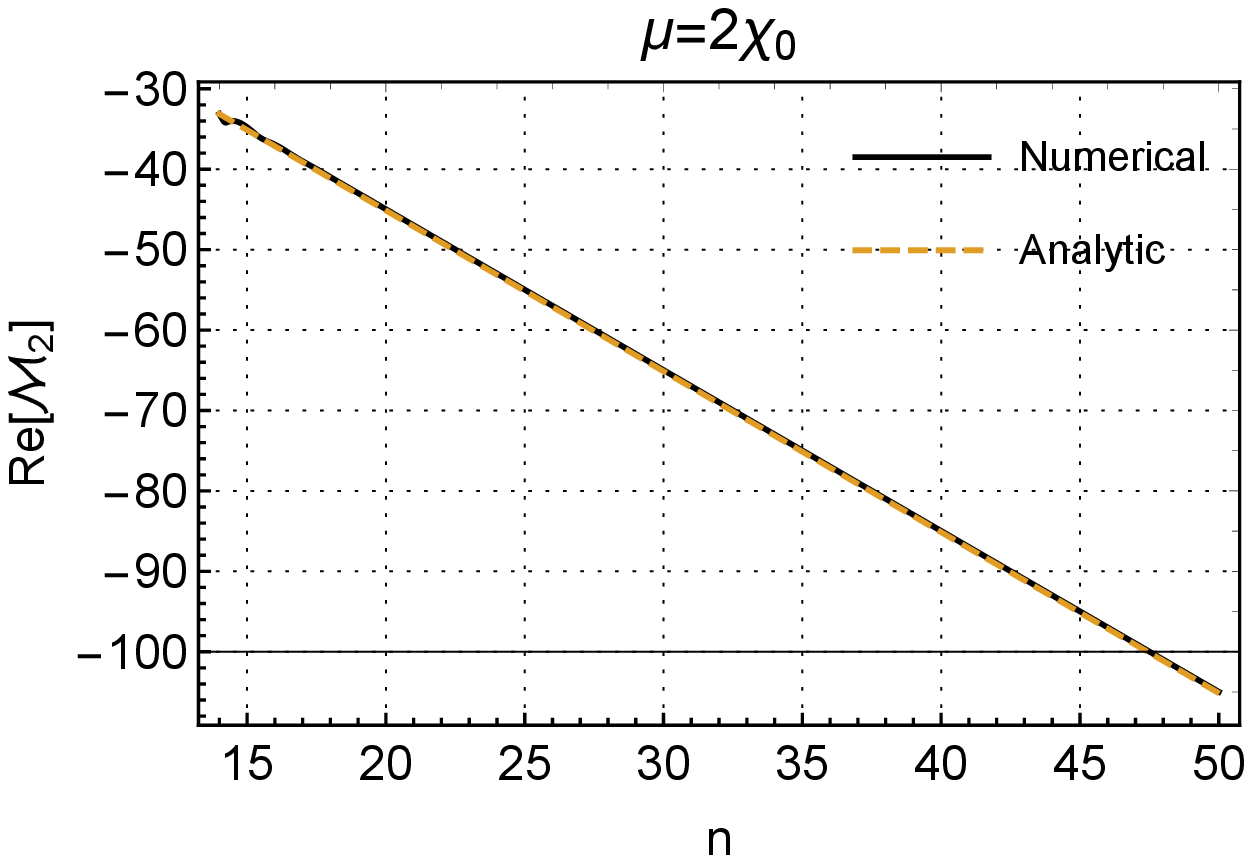}
		\caption{$\mu = 2 \chi_0$}
	\end{subfigure}
	\begin{subfigure}[b]{0.33\textwidth}
		\centering
		\includegraphics[width=\textwidth]{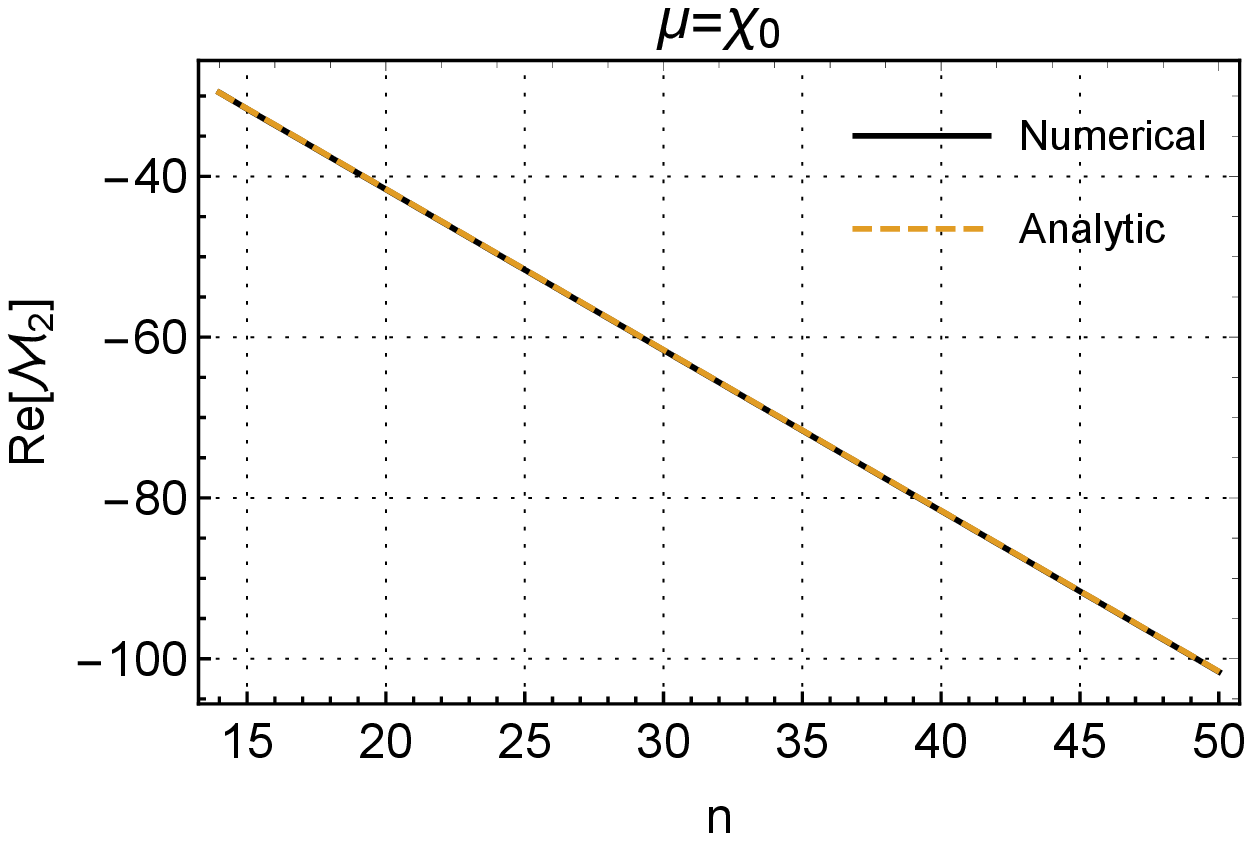}
		\caption{$\mu = \chi_0$}
	\end{subfigure}
	\begin{subfigure}[b]{0.33\textwidth}
		\centering
		\includegraphics[width=\textwidth]{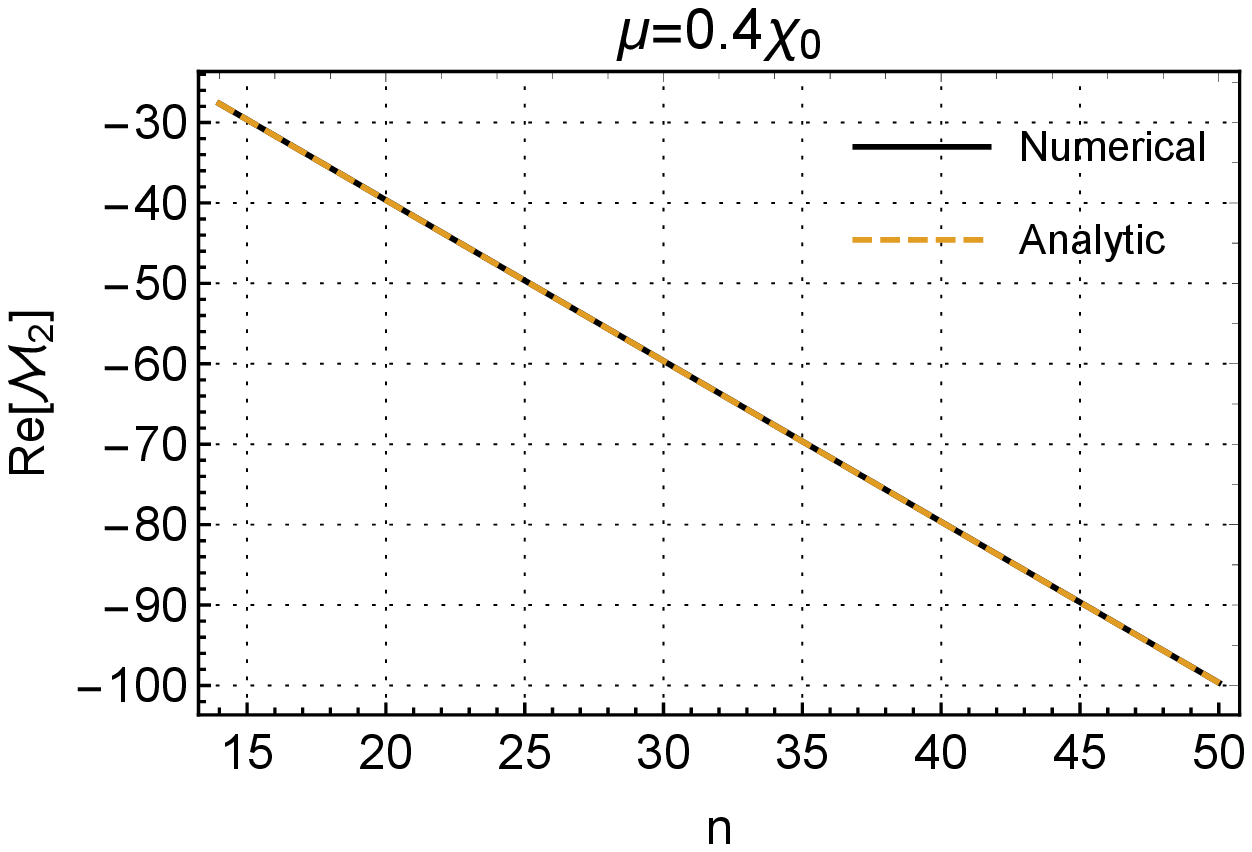}
		\caption{$\mu = \frac25 \chi_0$}
	\end{subfigure}%
	\caption{\footnotesize Comparing the real parts of $\mathcal{M}_{-}(n,3800 \chi_0,\mu)$ 
	and $\mathcal{M}_{-2}(n,3800 \chi_0,\mu)$ for $\mu = 2 \chi_0$ (left), $\mu = 
	\chi_0$ (middle) and $\mu = \frac25 \chi_0$ (right). In each case the numerical
	solution is solid blue while the approximation is long dashed yellow.}
	\label{M2real-figure}
\end{figure}

\begin{figure}[H]
	\centering
	\begin{subfigure}[b]{0.33\textwidth}
		\centering
		\includegraphics[width=\textwidth]{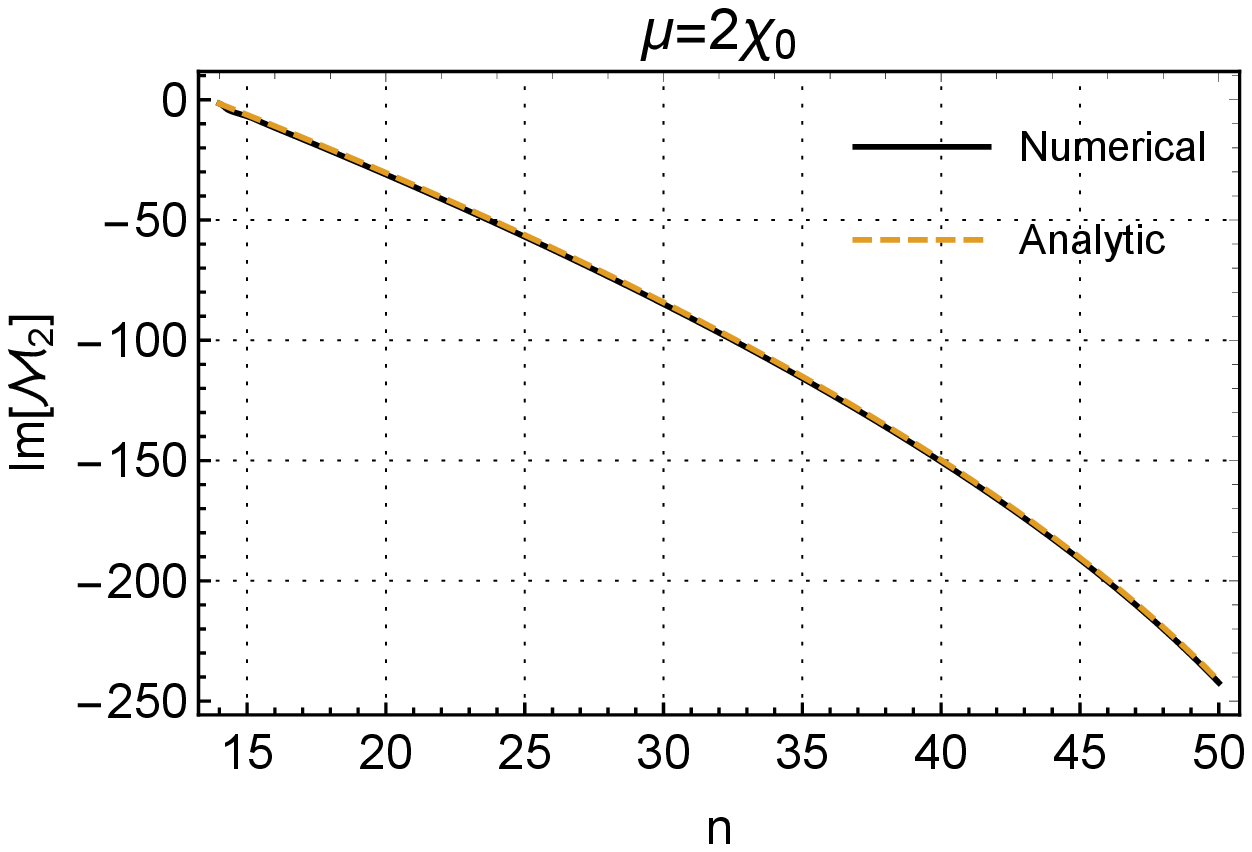}
		\caption{$\mu = 2 \chi_0$}
	\end{subfigure}
	\begin{subfigure}[b]{0.33\textwidth}
		\centering
		\includegraphics[width=\textwidth]{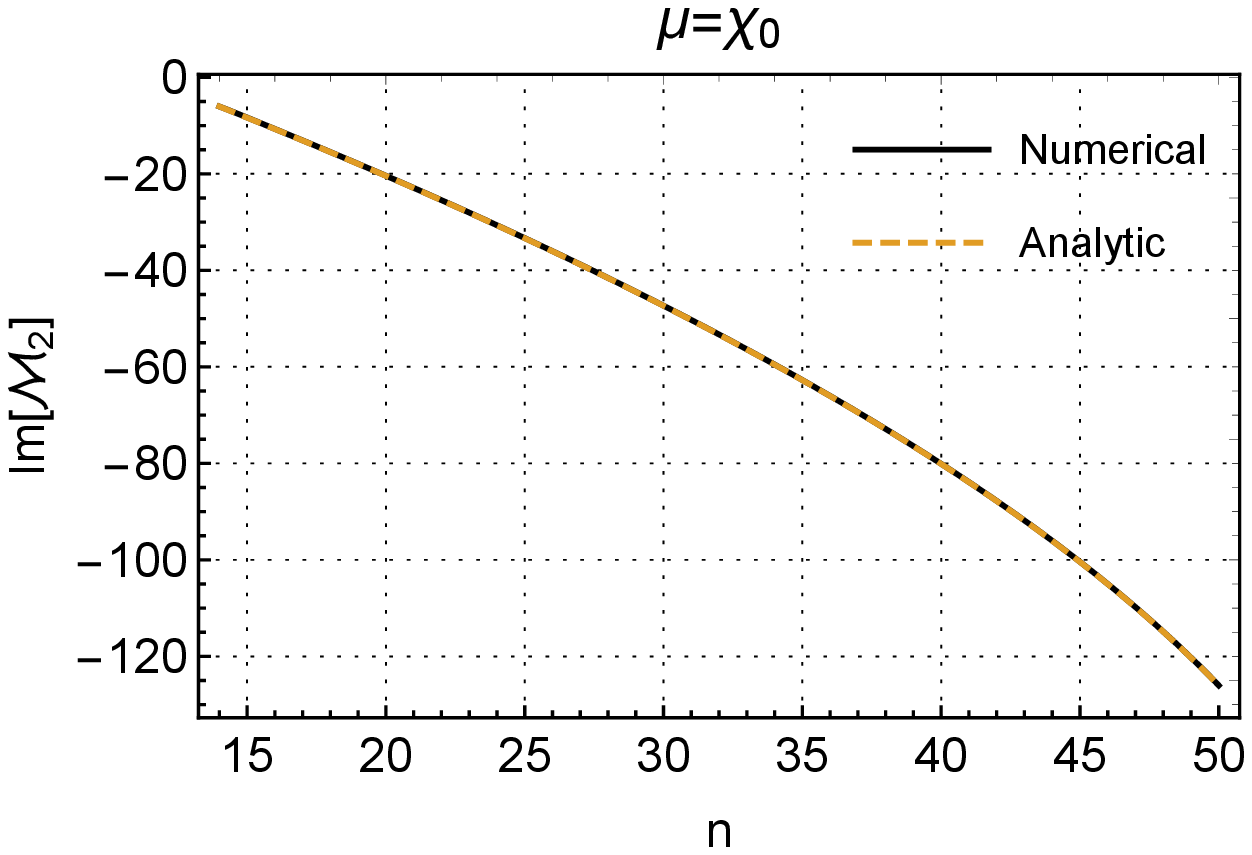}
		\caption{$\mu = \chi_0$}
	\end{subfigure}
	\begin{subfigure}[b]{0.33\textwidth}
		\centering
		\includegraphics[width=\textwidth]{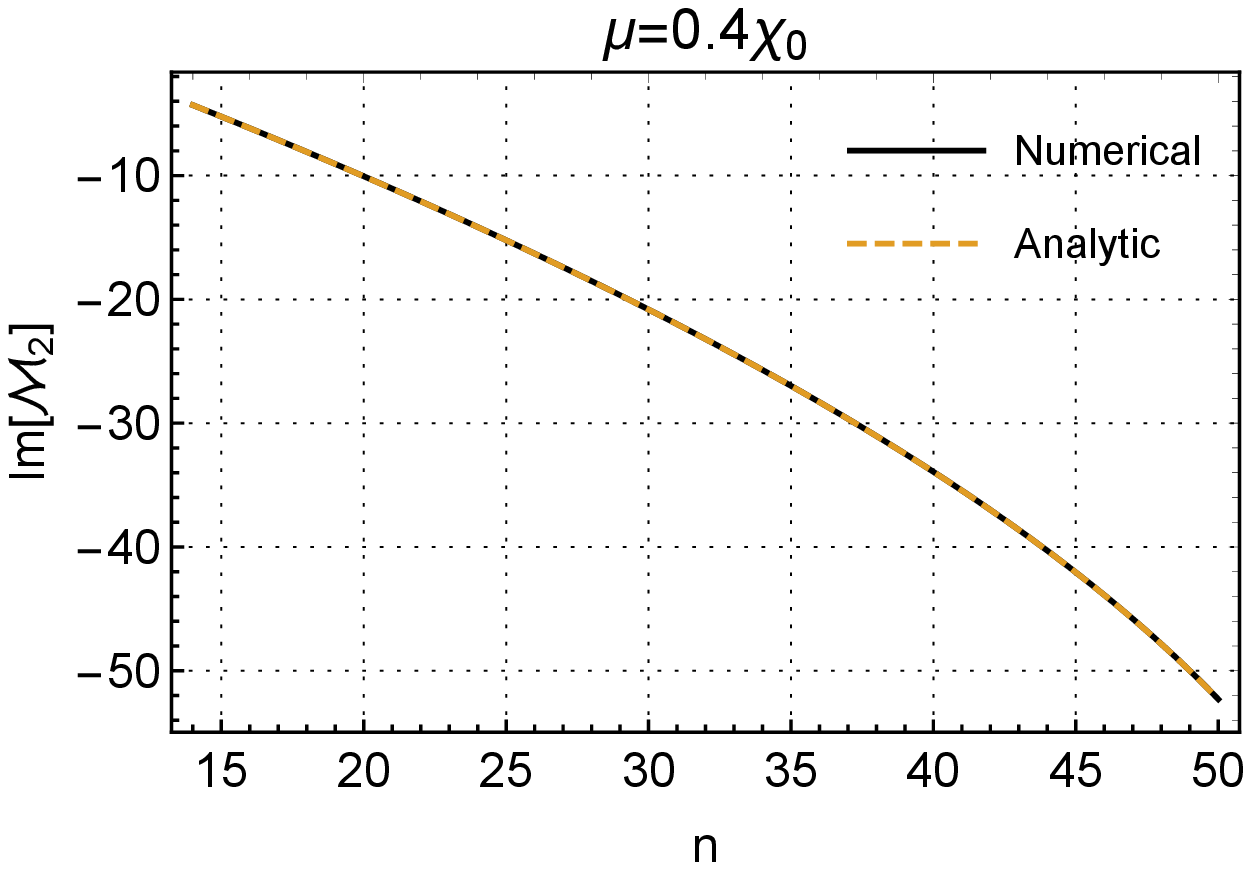}
		\caption{$\mu = \frac25 \chi_0$}
	\end{subfigure}%
	\caption{\footnotesize Comparing the imaginary parts of $\mathcal{M}_{-}(n,3800 
	\chi_0,\mu)$ and $\mathcal{M}_{-2}(n,3800 \chi_0,\mu)$ for $\mu = 2 \chi_0$ (left), 
	$\mu = \chi_0$ (middle) and $\mu = \frac25 \chi_0$ (right). In each case the numerical
	solution is solid blue while the approximation is long dashed yellow.}
	\label{M2imaginary-figure}
\end{figure}

It remains only to choose the point $n = n_2$ for making the transition from the 
$\mathcal{M}_{\pm}$ approximation to the post-horizon crossing approximation 
(\ref{rightsol}). Based on Figures \ref{M1real-figure} and \ref{M1imaginary-figure} 
it seems quite accurate to take $n_2 = n_{\kappa} + 4$. Hence we define the 
$\mathcal{M}_{\pm 2}$ approximation as,
\begin{equation}
\mathcal{M}_{\pm 2}(n,\kappa,\mu) = \mathcal{M}_{\pm 1}(n_2,\kappa,\mu) - 
2 (n \!-\! n_2) \mp 2 i \!\! \int_{n_2}^{n} \!\! \frac{\mu dn'}{\chi(n')} \; .
\label{M2def}
\end{equation}
Note that only the integration constant $\mathcal{M}_{\pm}(n_2,\kappa,\mu)$
depends on the dimensionless wave number $\kappa$. Figures \ref{M2real-figure} 
and \ref{M2imaginary-figure} compare the real and imaginary parts of this 
approximation with the numerical evolution. 
\noindent Agreement is excellent, not only for the real parts --- which roughly 
coincide with the $\mathcal{M}_{\pm 1}$ approximation in Figure \ref{M1real-figure} 
--- but also for the imaginary parts --- which show large deviations from the 
$\mathcal{M}_{\pm 1}$ approximation in Figure \ref{M1imaginary-figure}.

\subsection{Plateau Potentials}

The quadratic dimensionless potential $U(\phi) = \frac12 c^2 \phi^2$ was chosen for 
our detailed studies because the slow roll approximations (\ref{slowroll}) give simple, 
analytic expressions for the dimensionless Hubble parameter $\chi(n)$ and the first 
slow roll parameter $\epsilon(n)$. With the choice of $c \simeq 7.1 \times 10^{-6}$ this 
model is consistent with observations of the scalar amplitude and the scalar spectral 
index \cite{Aghanim:2018eyx}. However, the model is excluded by its high prediction of 
$r \simeq 0.14$ for the tensor-to-scalar ratio \cite{Aghanim:2018eyx}. It is worth
briefly considering how our analysis applies to the plateau potentials that are 
currently permitted by the data. 

\begin{figure}[H]
\includegraphics[width=4.5cm,height=4cm]{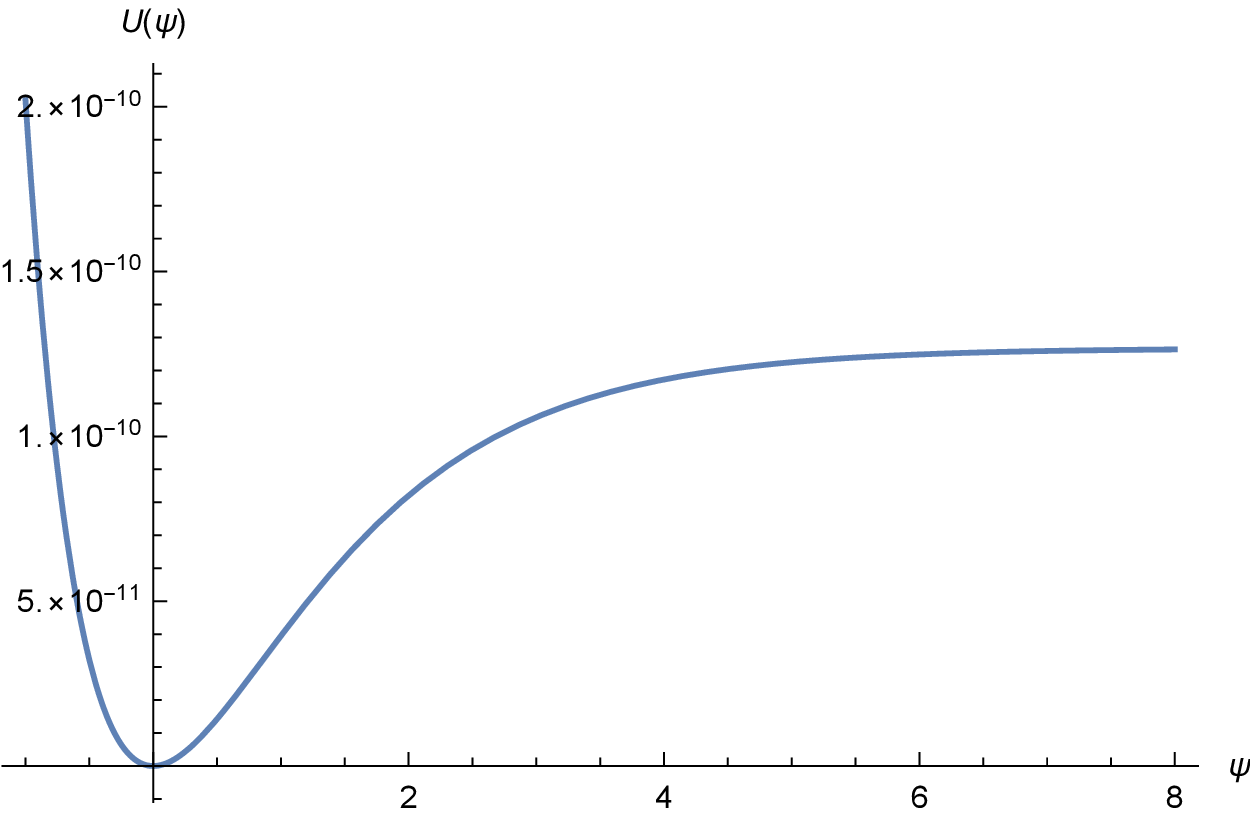}
\hspace{-0.1cm}
\includegraphics[width=4.5cm,height=4cm]{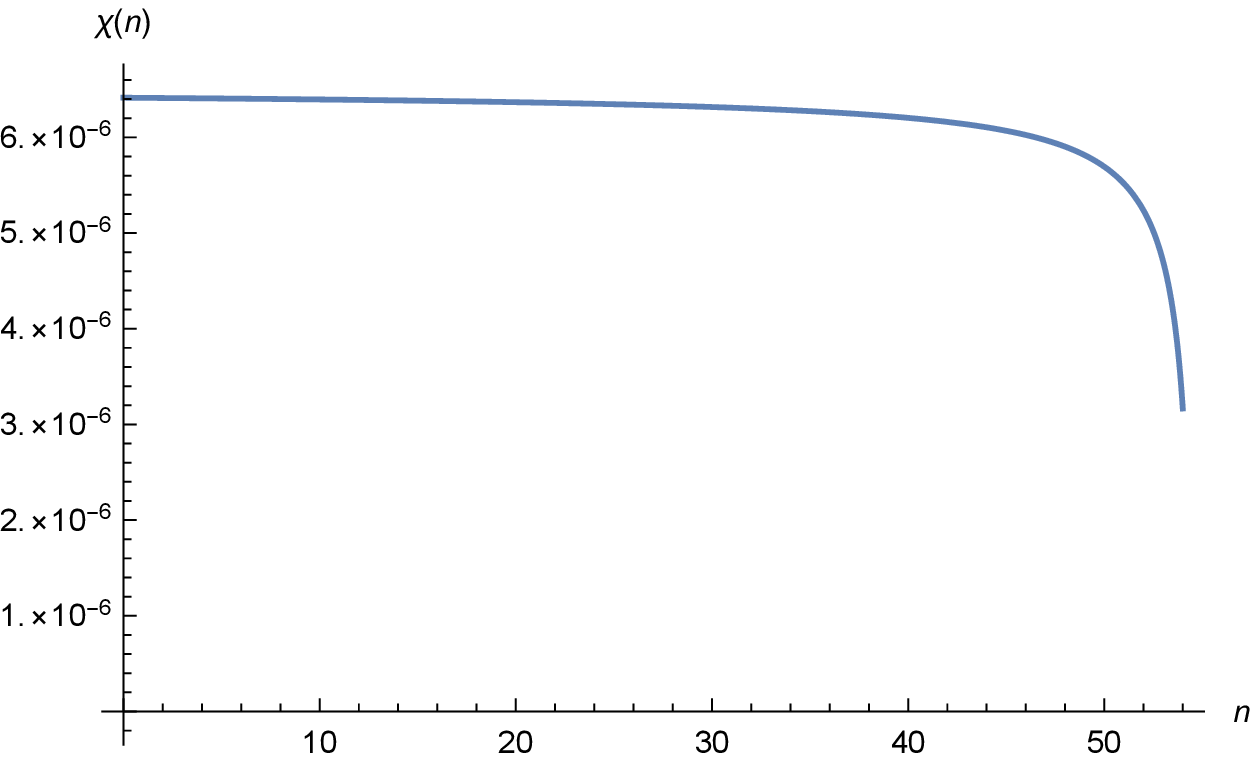}
\hspace{-0.1cm}
\includegraphics[width=4.5cm,height=4cm]{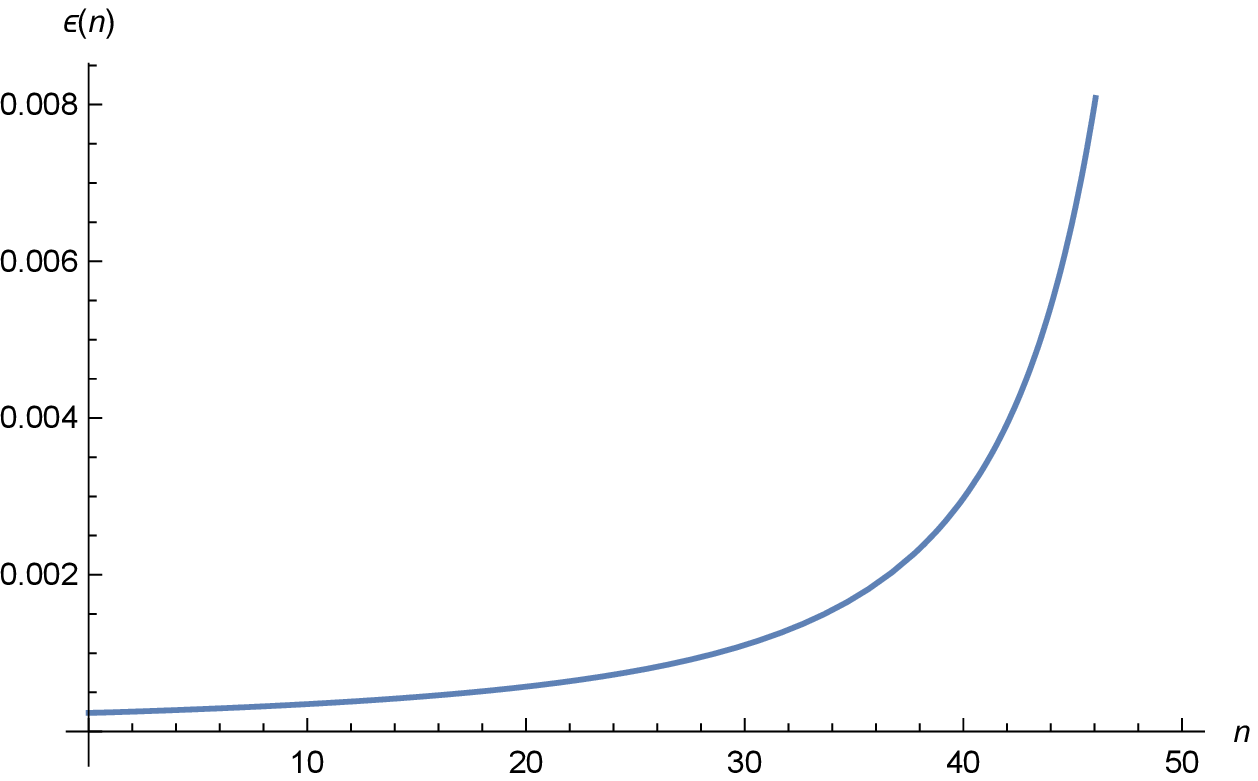}
\caption{\footnotesize{These graphs concern the Einstein-frame representation of
Starobinsky's original model of inflation \cite{Starobinsky:1980te}. The left hand
plot depicts the potential $U(\phi)$ (\ref{StaroU}); the middle graph gives the
dimensionless Hubble parameter $\chi(n)$ and the right hand plot shows the first
slow roll parameter $\epsilon(n)$. The geometrical quantities are associated with
starting inflation from $\phi_0 = 5.3$.}}
\label{Starobinsky1-figure}
\end{figure}

One of the simplest plateau potentials is the Einstein-frame version of Starobinsky's
famous $R + R^2$ model \cite{Starobinsky:1980te}. In our notation, the dimensionless 
potential is \cite{Brooker:2016oqa},
\begin{equation}
U(\phi) = \frac34 M^2 \Bigl( 1 - e^{-\sqrt{\frac23} \, \phi}\Bigr)^2 \qquad , \qquad
 M = 1.3 \times 10^{-5} \; . \label{StaroU}
\end{equation}
Starting from $\phi_0 = 5.3$ produces a little over 50 e-foldings of inflation, and the
model is not only consistent with observations of the scalar amplitude and spectral
but also with the upper limit on the tensor-to-scalar ratio \cite{Aghanim:2018eyx}. 
A glance at Figure~\ref{Starobinsky1-figure} reveals why $r = 16 \epsilon$ is so small: 
the dimensionless Hubble parameter $\chi(n)$ is almost constant.  

\begin{figure}[H]
	\centering
	\begin{subfigure}[b]{0.33\textwidth}
		\centering
		\includegraphics[width=\textwidth]{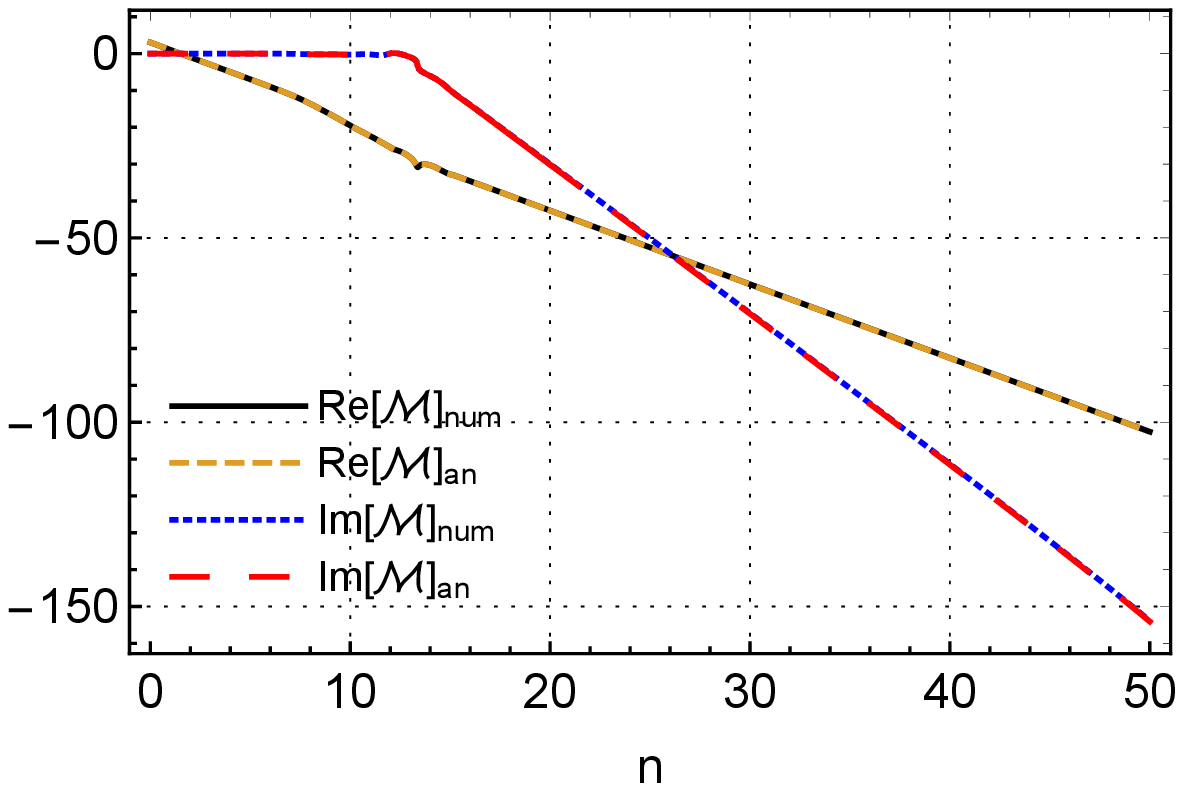}
		\caption{$\mu = 2 \chi_0$}
	\end{subfigure}
	\begin{subfigure}[b]{0.33\textwidth}
		\centering
		\includegraphics[width=\textwidth]{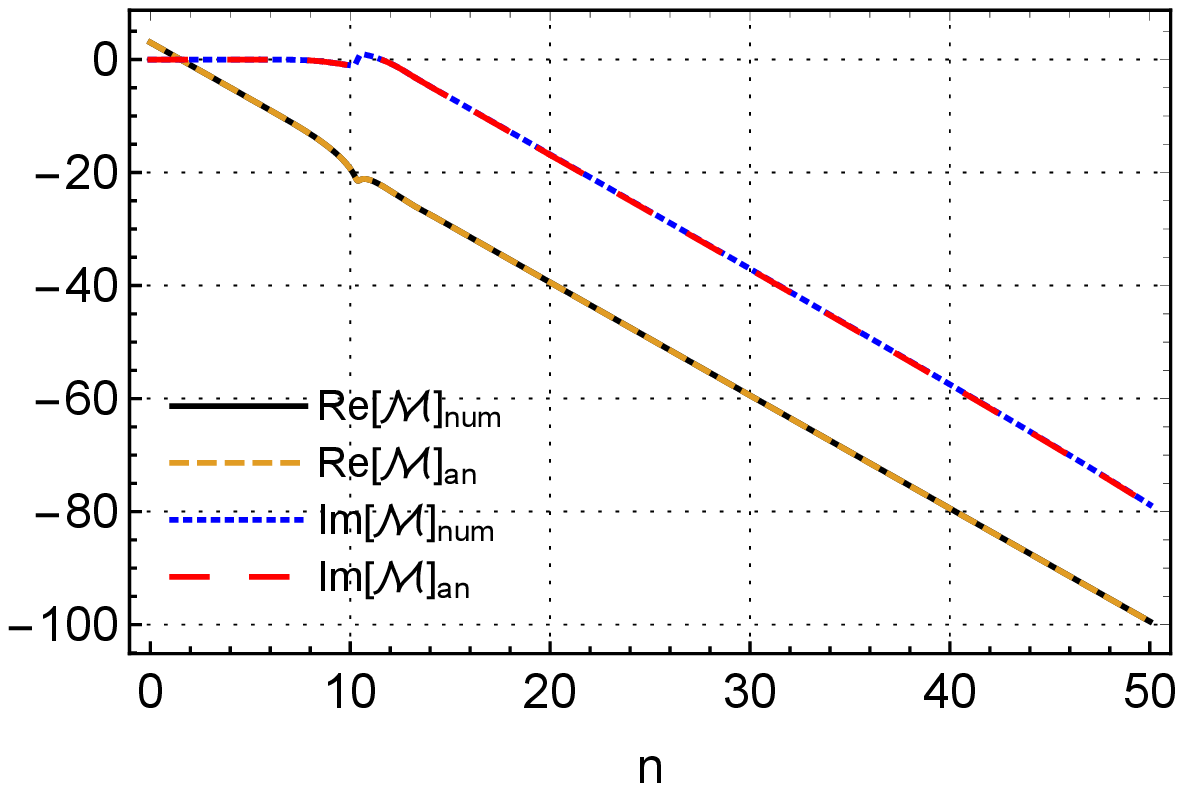}
		\caption{$\mu = \chi_0$}
	\end{subfigure}
	\begin{subfigure}[b]{0.33\textwidth}
		\centering
		\includegraphics[width=\textwidth]{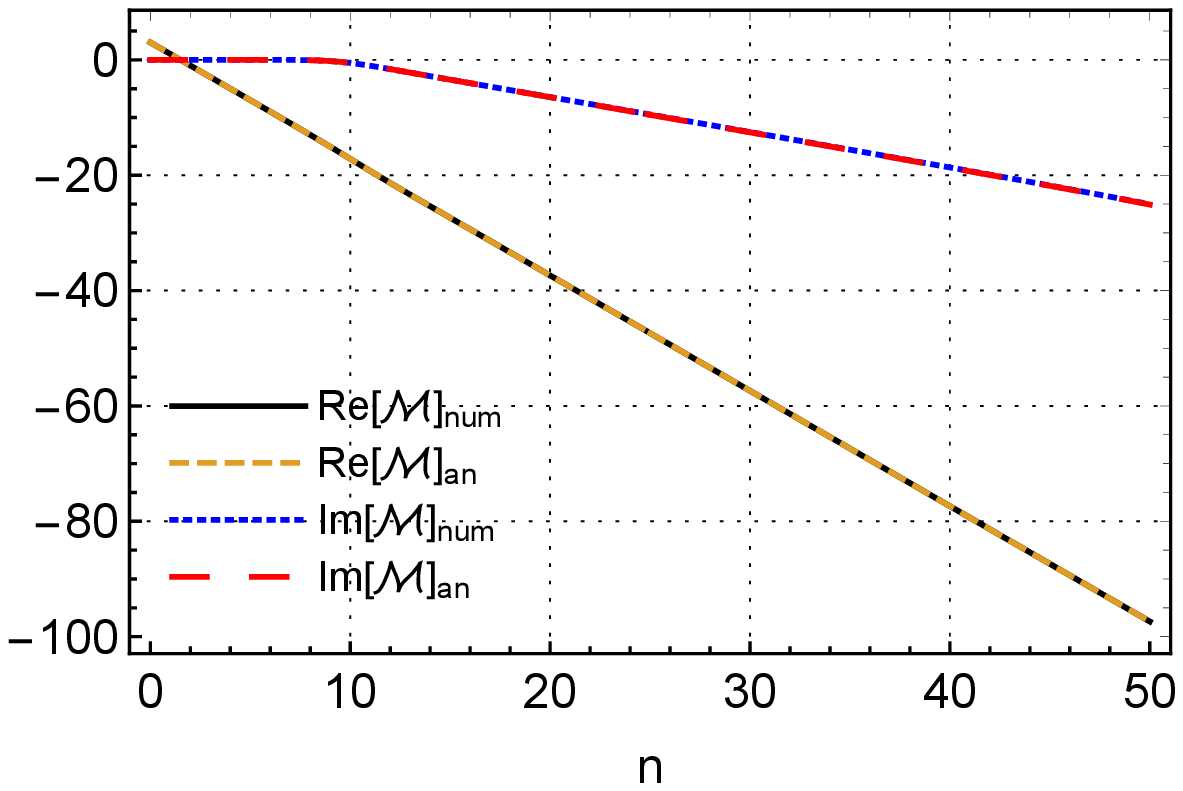}
		\caption{$\mu = \frac3{10} \chi_0$}
	\end{subfigure}%
	\caption{\footnotesize Comparison of the numerical $M_{-}(n,\kappa,\mu)$
	with our approximations (\ref{M1def}), for $0 < n < n_2$, and (\ref{M2def}),
	for $n_2 < n < 50$. The dimensionless wave number is $\kappa = 3800 \chi_0$
	and the dimensionless mass parameter is $\mu = 2 \chi_0$ (left), $\mu = \chi_0$
	(middle) and $\mu = \frac3{10} \chi_0$ (right). In each case the real part of 
	the numerical solution is in solid black while its imaginary part is in short
	dashed blue; the real part of the approximations are in long dashed yellow 
	while its imaginary part is in very long dashed red.}
	\label{Starobinsky2-figure}
\end{figure}

Our approximations (\ref{M1def}) and (\ref{M2def}) are independent of the classical
model, and Figure \ref{Starobinsky2-figure} demonstrates their validity for the
plateau potential (\ref{StaroU}). The chief difference between a plateau potential, 
and the quadratic model, is that the near constancy of $\chi(n)$ makes the imaginary 
part of the $M_{\pm 2}(n,\kappa,\mu)$ approximation (\ref{M2def}) nearly linear. 
This is apparent in Figure \ref{Starobinsky2-figure}, and contrasts with the 
curvature which is evident in Figure \ref{M2imaginary-figure}. However, for {\it 
both} potentials the approximations (\ref{M1def}) and (\ref{M2def}) are so good,
in the ranges of validity, that one cannot even discern a difference with the 
numerical result.

\section{The Inflaton Effective Potential}

The purpose of this section is to derive the one loop correction to the
inflaton effective potential. We begin by computing the primitive 
contribution from the dimensionally regulated trace of the fermion 
propagator. This is then renormalized and the unregulated limit is taken. 
The section closes by checking the de Sitter and flat space limits, and by
giving the large field and small field expansions.

\subsection{The Primitive Contribution}

Recall that the derivative of the effective potential with respect to
$\varphi$ is defined in terms of the trace of the coincident fermion 
propagator by equation (\ref{DVfermion}). The trace of the coincident
fermion propagator is the primitive contribution. Equation 
(\ref{fermiontrace}) gives it in terms of the coincidence limit of scalar 
propagators $i\Delta[\xi_c,M^2_{\pm}](x;x')$, where $\xi_c = \frac14 
(\frac{D-2}{D-1})$ and $M^2_{\pm} = f\varphi (f\varphi \mp iH) \equiv \mu 
(\mu \mp i \chi)/8\pi G$. Finally, we can use expression 
(\ref{scalar prop mode int}) to compute the coincidence limit of these 
scalar propagators,
\begin{equation}
i \Delta[\xi_c,M^2_{\pm}](x;x) = \frac{2 (8\pi G)^{\frac{2-D}{2}}}{
(4\pi)^{\frac{D-1}2} \Gamma(\frac{D-1}2)} \int_{0}^{\infty} \!\! d\kappa 
\, \kappa^{D-2} e^{\mathcal{M}_{\pm}(n,\kappa,\mu)} \; . \label{modesum}
\end{equation}

Recall from section 2 that we approximate $\mathcal{M}_{\pm}(n,\kappa,\mu)$
with expression (\ref{M1def}) for $0 \leq n \leq n_2 \equiv n_{\kappa} = 4$
and by expression (\ref{M2def}) for $n_2 \leq n$. These conditions must be 
translated from the e-folding $n$ to the dimensionless wave number $\kappa$ 
in order to apply them the integration (\ref{modesum}). To make this 
translation note that just as each $\kappa$ has an e-folding $n_{\kappa}$ at 
which it experiences horizon crossing ($\kappa = e^{n_{\kappa}} \chi(n_{\kappa})$),
provided inflation lasts long enough, so too we can regard each e-folding $n$ as 
having a wave number $\kappa_{n}$ at which $\kappa_n = e^{n} \chi(n)$. Hence
the cross-over between (\ref{M1def}) and (\ref{M2def}) corresponds to $\kappa = 
\kappa_{n-4}$. Because only the large $\kappa$ portion of the integration 
requires dimensional regularization we have,
\begin{eqnarray}
\lefteqn{i\Delta[\xi_c,M^2_{\pm}](x;x) \simeq \frac{2 (8\pi G)^{\frac{2-D}{2}}}{
(4\pi)^{\frac{D-1}2} \Gamma(\frac{D-1}2)} \int_{\kappa_{n-4}}^{\infty} \!\!\!\!\! 
d\kappa \, \kappa^{D-2} e^{\mathcal{M}_{\pm 1}(n,\kappa,\mu)}  } \nonumber \\
& & \hspace{6.5cm} +  \frac1{16 \pi^3 G} \! \int_{0}^{\kappa_{n-4}} \!\!\!\!\! 
d\kappa \, \kappa^2 e^{\mathcal{M}_{\pm 2}(n,\kappa,\mu)} . \qquad \label{modesum2}
\end{eqnarray}
By extending the range of integration for the $\mathcal{M}_{\pm 1}$ approximation 
all the way to $\kappa = 0$, and subtracting the extension from the second line of
(\ref{modesum2}), we at length reach the form,
\begin{equation}
i\Delta[\xi_c,M^2_{\pm}](x;x) \simeq i\Delta_{\pm 1}(x;x) + \! \int_{0}^{\kappa_{n-4}} 
\!\!\! \frac{d\kappa \, \kappa^2}{16 \pi^3 G} \Bigl[e^{\mathcal{M}_{\pm 2}(n,\kappa,\mu)} 
- e^{\mathcal{M}_{\pm 1}(n,\kappa,\mu)}\Bigr] , \label{modesum3}
\end{equation}
where $i\Delta_{\pm 1}(x;x')$ is the propagator defined by the Hankel
functions of the $\mathcal{M}_{\pm 1}$ approximation. The coincidence limit of
this propagator can be evaluated using integral $6.574\; \# 2$ of 
\cite{Gradshteyn:1965},
\begin{equation}
i\Delta_{\pm 1}(x;x) = \frac{[(1 \!-\! \epsilon) H]^{D-2}}{
(4\pi)^{\frac{D}2}} \times \frac{\Gamma( \frac{D-1}{2} \!+\! \nu_{\pm}) 
\Gamma(\frac{D-1}{2} \!-\! \nu_{\pm})}{\Gamma(\frac12 \!+\! \nu_{\pm}) 
\Gamma(\frac12 \!-\! \nu_{\pm} )} \times \Gamma\Bigl(1 \!-\! \frac{D}2\Bigr) 
\; , \label{delta1 integral}
\end{equation}
where $\nu^2_{\pm}$ was defined in expression (\ref{hankel index}).

\subsection{Renormalization}

Recall that the derivative of the effective potential (\ref{DVfermion}) was 
expressed in equation (\ref{fermiontrace}) using the same coincident scalar 
propagators we have just approximated in expression (\ref{modesum3}). What we 
might call the $\Delta V_{1}'$ approximation arises from using just
$i\Delta_{\pm 1}(x;x)$ in equations (\ref{DVfermion}) and (\ref{fermiontrace}),
\begin{eqnarray}
\lefteqn{\Delta V_{1}' = \delta \xi \varphi R + \frac16 \delta \lambda 
\varphi^3 - f \Bigl[ 2m \!+\! i H \Bigl(D \!-\! 2 + \frac{d}{dn}\Bigr) \Bigr] 
i \Delta_{+ 1}(x;x) } \nonumber \\
& & \hspace{5cm} - f \Bigl[ 2m \!-\! i H \Bigl(D \!-\! 2 + \frac{d}{dn}\Bigr) 
\Bigr] i \Delta_{- 1}(x;x) \; . \qquad \label{DV1A}
\end{eqnarray}
Note that $\Delta V_1'$ is real even though the $i\Delta_{\pm1}(x;x)$ are
complex. Now expand (\ref{delta1 integral}) in powers of $D - 4$,
\begin{eqnarray}
\lefteqn{ i\Delta_{\pm 1}(x;x) = \frac{[(1 \!-\! \epsilon) H]^{D-2}}{
(4\pi)^{\frac{D}2}} \Bigl[ \Bigl( \frac{D \!-\! 3}{2}\Bigr)^2 - \nu^2_{\pm}\Bigr]}
\nonumber \\
& & \hspace{2.2cm} \times \Biggl\{ \Gamma\Bigl(1 \!-\! \frac{D}2\Bigr) +
\psi\Bigl(\frac12 \!+\! \nu_{\pm} \Bigr) + \psi\Bigl(\frac12 \!-\! \nu_{\pm} 
\Bigr) + O(D \!-\! 4) \Biggr\} , \qquad \label{Delta1B} \\
& & = \frac{[(1 \!-\! \epsilon) H]^{D-4}}{(4\pi)^{\frac{D}2}} \Biggl\{ 
\Gamma\Bigl(1 \!-\! \frac{D}2\Bigr) M^2_{\pm} + (1 \!-\! \epsilon)^2 H^2 
\nonumber \\
& & \hspace{3.9cm} + \Bigl[\psi\Bigl(\frac12 \!+\! \nu_{\pm} \Bigr) + 
\psi\Bigl(\frac12 \!-\! \nu_{\pm} \Bigr) \Bigr] M^2_{\pm} + O(D \!-\! 4) 
\Biggr\} , \qquad \label{Delta1C}
\end{eqnarray}
where $\psi(x) \equiv \frac{d}{dx} \ln[\Gamma(x)]$ is the digamma function and 
we recall that $M^2_{\pm} \equiv f \varphi [f \varphi \mp i H]$. Note also
that expression (\ref{hankel index}) suggests a very simple approximation
for the index that can be used for finite terms,
\begin{equation}
\nu^2_{\pm} = \Bigl[ \frac12 \pm \frac{i f\varphi}{(1 \!-\! \epsilon) H}\Bigr]^2
\pm \frac{i \epsilon f \varphi}{(1 \!-\! \epsilon)^2 H} \qquad \Longrightarrow
\qquad \nu_{\pm} \simeq \frac12 \pm \frac{i f\varphi}{(1 \!-\! \epsilon) H} 
\; . \label{nuapprox}
\end{equation}

The next step is to substitute each of the three terms from (\ref{Delta1C})
into expression (\ref{DV1A}). The only divergences are associated with the 
term proportional to $\Gamma(1 - \frac{D}2) = \frac{2}{D-4} + O(1)$. Including 
the two counterterms gives,
\begin{eqnarray}
\lefteqn{ \Bigl(\Delta V_{1}'\Bigr)_{\rm 1st} = \delta \xi \varphi R + 
\frac16 \delta \lambda \varphi^3 + \frac{[(1 \!-\! \epsilon) H]^{D-4}}{
(4\pi)^{\frac{D}2}} \Biggl\{ -\Gamma\Bigl(1 \!-\! \frac{D}2\Bigr) 
\Bigl[4 f^4 \varphi^3 + \frac{f^2 \varphi R}{D \!-\! 1} \Bigr] } \nonumber \\
& & \hspace{5cm} + \Bigl[-1 \!+\! 2 \epsilon \!+\! \frac{2 \epsilon'}{1 \!-\! 
\epsilon}\Bigr] 2 f^2 \varphi H^2 + O(D \!-\! 4) \Biggr\} . \qquad \label{DV1B}
\end{eqnarray}
We choose the counterterms to cancel the divergences,
\begin{equation}
\delta\xi = \frac{f^2 s^{D-4} \Gamma(1 \!-\! \frac{D}2)}{(D \!-\! 1) 
(4\pi)^{\frac{D}{2}}} \qquad , \qquad \delta \lambda = 
\frac{24 f^4 s^{D-4} \Gamma(1 \!-\! \frac{D}2)}{(4\pi)^{\frac{D}{2}}} \; ,
\label{cterms}
\end{equation}
where $s$ is the dimensional regularization scale. Note that the divergent
parts of our counterterms agree with those of de Sitter background (equations 
(51-52) of \cite{Miao:2006pn}, and equations (16-17) of \cite{Miao:2015oba}). 
This is as it should be because counterterms are background-independent. 
Taking the unregulated limit of (\ref{DV1B}) with these counterterms and
integrating gives,
\begin{equation}
\Bigl(\Delta V_{1} \Bigr)_{\rm 1st} \longrightarrow -\frac{(\frac16 f^2 
\varphi^2 R \!+\! f^4 \varphi^4)}{8 \pi^2} \ln\Bigl[ \frac{(1 \!-\! \epsilon) 
H}{s}\Bigr] - \frac{f^2 \varphi^2 H^2}{8 \pi^2} \Bigl[\frac12 \!-\! \epsilon 
\!-\! \frac{\epsilon'}{1 \!-\! \epsilon}\Bigr] . \label{V1A}
\end{equation}

The second term in (\ref{Delta1C}) is purely real so it makes a simple 
contribution,
\begin{equation}
\Bigl( \Delta V_{1}'\Bigr)_{\rm 2nd} -\frac{2 f^2 \varphi (1 \!-\! \epsilon)^2
H^2}{8 \pi^2} \qquad \Longrightarrow \qquad \Bigl( \Delta V_{1}\Bigr)_{\rm 2nd}
= -\frac{f^2 \varphi^2 (1 \!-\! \epsilon)^2 H^2}{8 \pi^2} \; .
\label{V1B}
\end{equation}
The most complicated contribution comes from the 3rd term of (\ref{Delta1C}),
which involves the digamma functions,
\begin{eqnarray}
\lefteqn{ \Bigl( \Delta V_{1}'\Bigr)_{\rm 3rd} = - \frac{(2 f^4 \varphi^3 \!+\! 
\frac16 f^2 \varphi R \!+\! f^2 \varphi H^2 \frac{d}{d n})}{8 \pi^2} 
{\rm Re}\Biggl[\psi\Bigl( \frac12 \!+\! \nu_{+}\Bigr) + \psi\Bigl( \frac12 \!-\! 
\nu_{+}\Bigr) \Biggr] } \nonumber \\
& & \hspace{4.5cm} + \frac{f^3 \varphi^2 H \frac{d}{d n}}{8 \pi^2} 
{\rm Im}\Biggl[\psi\Bigl( \frac12 \!+\! \nu_{+}\Bigr) + \psi\Bigl( \frac12 \!-\! 
\nu_{+}\Bigr)\Biggr] . \qquad \label{DV1D}
\end{eqnarray}
Integrating gives,
\begin{eqnarray}
\lefteqn{ \hspace{-0.2cm} \Bigl( \Delta V_{1}\Bigr)_{\rm 3rd} \!\!\!= -
\frac{H^4}{8 \pi^2} \int_{0}^{\frac{f\varphi}{H}} \!\!\!\!\!\! dx \Bigl[ 
2 x + 2 x^3\Bigr] {\rm Re}\Biggl[\psi\Bigl( \frac12 \!+\! \nu(x)\Bigr) + 
\psi\Bigl( \frac12 \!-\! \nu(x)\Bigr) \Biggr] } \nonumber \\
& & \hspace{0.5cm} - \frac{H^4}{8 \pi^2} \Bigl[ \frac{d}{d n} - 3 \epsilon\Bigr] 
\!\! \int_{0}^{\frac{f\varphi}{H}} \!\!\!\!\!\! dx \Biggl\{ x {\rm Re}\Biggl[
\psi\Bigl( \frac12 \!+\! \nu(x)\Bigr) + \psi\Bigl( \frac12 \!-\! \nu(x)\Bigr) 
\Biggr] \nonumber \\
& & \hspace{4.5cm} - x^2 {\rm Im}\Biggl[\psi\Bigl( \frac12 \!+\! \nu(x)\Bigr) + 
\psi\Bigl( \frac12 \!-\! \nu(x)\Bigr) \Biggr] \Biggr\} , \qquad \label{V1C}
\end{eqnarray}
where $\nu(x) \equiv \sqrt{\frac14 - \frac{(x^2 - i x)}{(1 - \epsilon)^2}} 
\simeq \frac12 + \frac{i x}{1 - \epsilon}$. Combining expressions (\ref{V1A}),
(\ref{V1B}) and (\ref{V1C}) gives a final form for the local part of the effective
potential,
\begin{equation}
\Delta V_{1} = -\frac{H^4}{8 \pi^2} \Biggl\{ F\Bigl( \frac{f \varphi}{H},
\epsilon\Bigr) + \Bigl[ (2 \!-\! \epsilon) \Bigl( \frac{f \varphi}{H}\Bigr)^2
+ \Bigl( \frac{f \varphi}{H}\Bigr)^4\Bigr] \ln\Bigl[ \frac{(1 \!-\! \epsilon) 
H}{s}\Bigr] \Biggr\} , \label{one loop correction}
\end{equation}
where the function $F(z,\epsilon)$ is,
\begin{eqnarray}
\lefteqn{ F(z,\epsilon) \equiv \Bigl[ 1 - 2 \epsilon - \frac{2 \epsilon'}{1 \!-\! 
\epsilon}\Bigr] z^2 + (1 \!-\! \epsilon)^2 z^2 } \nonumber \\
& & + \int_{0}^{z} \!\!\! dx \Bigl[ 2 x + 2 x^3\Bigr] 
{\rm Re}\Biggl[\psi\Bigl( \frac12 \!+\! \nu(x)\Bigr) + \psi\Bigl( \frac12 \!-\! 
\nu(x)\Bigr) \Biggr] \nonumber \\
& & \hspace{1cm} + \Bigl[ \frac{d}{d n} - 3 \epsilon\Bigr] 
\!\! \int_{0}^{z} \!\!\! dx \Biggl\{ x {\rm Re}\Biggl[
\psi\Bigl( \frac12 \!+\! \nu(x)\Bigr) + \psi\Bigl( \frac12 \!-\! \nu(x)\Bigr) 
\Biggr] \nonumber \\
& & \hspace{4.5cm} - x^2 {\rm Im}\Biggl[\psi\Bigl( \frac12 \!+\! \nu(x)\Bigr) + 
\psi\Bigl( \frac12 \!-\! \nu(x)\Bigr) \Biggr] \Biggr\} . \qquad \label{Fdef}
\end{eqnarray}
Note that $F(z,\epsilon)$ is real in spite of the complex index 
$\nu(x) = \sqrt{\frac14 - \frac{(x^2 - i x)}{(1 - \epsilon)^2}}$.

\subsection{Correspondence Limits and Expansions}

We recover the de Sitter result by setting $\epsilon = 0$ which means $H$ is
constant,
\begin{eqnarray}
\lefteqn{\Delta V_{\rm dS}(\varphi) = -\frac{H^4}{8 \pi^2} \Biggl\{ 
\Bigl[ 2 \Bigl( \frac{f \varphi}{H}\Bigr)^2
+ \Bigl( \frac{f \varphi}{H}\Bigr)^4\Bigr] \ln\Bigl[ \frac{H}{s}\Bigr] +
2 \Bigl( \frac{f \varphi}{H}\Bigr)^2 } \nonumber \\
& & \hspace{4.4cm} + \int_{0}^{\frac{f\varphi}{H}} \!\!\!\!\! dx \, 
(2 x \!+\! 2 x^3) \Bigl[ \psi(1 \!+\! i x) + \psi(1 \!-\! i x)\Bigr] \Biggr\} 
. \qquad \label{DV desitter limit}
\end{eqnarray}
This agrees with the result of Candelas and Raine \cite{Candelas:1975du},
up to finite renormalizations of the $\varphi^2$ and $\varphi^4$ terms.

It will be seen that most of the terms in expression (\ref{one loop correction})
depend on the dimensionless ratio $f \varphi/H$. Hence the large field regime is 
the same as the small $H$ regime. We can access this regime by employing the 
large argument expansion for the digamma function in expression (\ref{Fdef}),
\begin{equation}
\vert x\vert \gg 1 \qquad \Longrightarrow \qquad \psi(x) = \ln(x) - \frac1{2 x} 
- \frac1{12 x^2} + \frac1{120 x^4} + O\Bigl(\frac1{x^6} \Bigr) \; . \qquad 
\label{largepsi}
\end{equation}
Applying (\ref{largepsi}) to the combination of digamma functions in (\ref{Fdef}) 
gives,
\begin{eqnarray}
\lefteqn{ \psi\Bigl( \frac12 \!+\! \nu(x)\Bigr) + \psi\Bigl( \frac12 \!-\! \nu(x)
\Bigr) = \ln\Bigl[ \frac14 \!-\! \nu^2\Bigr] - \frac{\frac{1}{3}}{\frac14 \!-\! \nu^2}
- \frac{\frac{1}{15}}{(\frac14 \!-\! \nu^2)^2} + \dots \; , } \\
& & \hspace{1cm} = \ln\Bigl[ \frac{x \sqrt{x^2 \!+\! 1}}{(1 \!-\! \epsilon^2)}\Bigr]
- \frac{(1 \!-\! \epsilon)^2}{3 (x^2 \!+\! 1)} - \frac{(1 \!-\! \epsilon)^4}{15 
(x^2 \!+\! 1)^2} + \dots \nonumber \\
& & \hspace{3cm} - i\Biggl[ {\rm tan}^{-1}\Bigl(\frac1{x}\Bigr) + \frac{(1 \!-\! 
\epsilon)^2}{3 x (x^2 \!+\! 1)} + \frac{2 (1 \!-\! \epsilon)^4}{15 x (x^2 \!+\! 1)^2} 
+ \dots \Biggr] \; . \qquad \label{psiexp}
\end{eqnarray}
Integrating term-by-term and making some additional expansions produces,
\begin{eqnarray}
\lefteqn{\Delta V_1 = - \frac{H^4}{8 \pi^2} \Biggl\{ \Bigl( \frac{f \varphi}{H}\Bigr)^4
\ln\Bigl( \frac{f \varphi}{s}\Bigr) - \frac14 \Bigl( \frac{f \varphi}{H}\Bigr)^4 
+ (2 \!-\! \epsilon) \Bigl( \frac{f \varphi}{H}\Bigr)^2 \ln\Bigl(\frac{f \varphi}{s}\Bigr)
} \nonumber \\
& & \hspace{-0.5cm} + \Bigl[ \frac12 \!-\! \epsilon \!+\! \frac23 (1 \!-\! \epsilon)^2 
\!-\! \frac{\epsilon'}{1 \!-\! \epsilon}\Bigr] \Bigl(\frac{f \varphi}{H}\Bigr)^2 
+ \Bigl[ \frac12 \!-\! \frac2{15} (1 \!-\! \epsilon)^4\Bigr] \ln\Bigl( 
\frac{f \varphi}{H}\Bigr) + O(1) \Biggr\} . \qquad \label{large field exp}
\end{eqnarray}
Of course the leading term is the famous result of Coleman and Weinberg 
\cite{Coleman:1973jx}. Note also that all the terms on the first line of 
(\ref{large field exp}) could be removed by allowed subtractions. 

To the small field regime we first expand the index $\nu(x)$,
\begin{eqnarray}
\nu(x) & \equiv & \sqrt{\frac14 - \frac{(x^2 \!-\! i x)}{(1 \!-\! \epsilon)^2}}
\equiv \frac12 - \Delta \nu \; , \\
& = & \frac12 - \frac{(x^2 \!-\! i x)}{(1 \!-\! \epsilon)^2} \Biggl\{ 1 +
\frac{(x^2 \!-\! i x)}{(1 \!-\! \epsilon)^2} + \frac{2 (x^2 \!-\! i x)^2}{ 
(1 \!-\! \epsilon)^4} + \dots \Biggr\} . \qquad
\end{eqnarray}
Then expand the digamma functions of (\ref{Fdef}) in powers of $\Delta \nu$,
\begin{eqnarray}
\lefteqn{\psi\Bigl( \frac12 \!+\! \nu\Bigr) + \psi\Bigl( \frac12 \!-\! \nu\Bigr)
= -\frac1{\Delta \nu} - 2 \gamma - 2 \sum_{n=1}^{\infty} \zeta(1 \!+\! 2 n) 
\Delta \nu^{2n} \; . } \\
& & \hspace{2cm} = \Bigl[ 1 - 2 \gamma - (1 \!-\! \epsilon)^2 + O(x^2)\Bigr]
+ i \Bigl[- \frac{(1 \!-\! \epsilon)^2}{x} + O(x) \Bigr] \; . \qquad 
\label{smallnuexp}
\end{eqnarray}
Substituting (\ref{smallnuexp}) in (\ref{Fdef}) and combining with 
(\ref{one loop correction}) gives,
\begin{equation}
\Delta V_1 = -\frac{H^4}{8 \pi^2} \Biggl\{ \Biggl[ \Bigl( 1 - \gamma
+ \ln\Bigl[ \frac{(1 \!-\! \epsilon) H}{s}\Bigr]\Bigr) (2 - \epsilon)
- \frac32 \epsilon - \frac{2 \epsilon'}{1 \!-\! \epsilon} \Biggr] \Bigl( 
\frac{f \varphi}{H}\Bigr)^2 + \dots \Biggr\} . \label{small field exp}
\end{equation}

\subsection{The Nonlocal Contribution}

We define the second integral of expression (\ref{modesum3}) as the infrared 
part of the propagator,
\begin{equation}
i\Delta_{\pm {\rm IR}}(x;x) \equiv \int_{0}^{\kappa_{n-4}} \!\!\! 
\frac{d\kappa \, \kappa^2}{16 \pi^3 G} \Bigl[e^{\mathcal{M}_{\pm 2}(n,\kappa,\mu)} 
- e^{\mathcal{M}_{\pm 1}(n,\kappa,\mu)}\Bigr] \; . \label{IRdef}
\end{equation}
By factoring out $\mathcal{M}_{\pm 1}(n_2,\kappa,\mu)$, and making the slow roll
approximation for the amplitude near horizon crossing we obtain,
\begin{eqnarray}
i\Delta_{\pm {\rm IR}}(x;x) & = & \int_{0}^{\kappa_{n-4}} \!\! 
\frac{d\kappa \, \kappa^2 e^{\mathcal{M}_{\pm}(n_2,\kappa,\mu)}}{16 \pi^3 G} 
\Bigl[e^{f_{\pm 2}(n,\kappa,\mu)} - e^{f_{\pm 1}(n,\kappa,\mu)}\Bigr] \; , 
\qquad \\
& \simeq & \int_{0}^{\kappa_{n-4}} \!\! \frac{d\kappa}{\kappa} 
\frac{\chi^2(n_{\kappa})}{32 \pi^3 G} \Bigl[e^{f_{\pm 2}(n,\kappa,\mu)} - 
e^{f_{\pm 1}(n,\kappa,\mu)}\Bigr] \; . \qquad
\end{eqnarray}
where we define,
\begin{eqnarray}
f_{\pm 1}(n,\kappa,\mu) & \!\!\!\! \equiv \!\!\!\! & 
\mathcal{M}_{\pm 1}(n,\kappa,\mu) - \mathcal{M}_{\pm 1}(n_2,\kappa,\mu) \; , \\
f_{\pm 2}(n,\kappa,\mu) & \!\!\!\! \equiv \!\!\!\! & 
\mathcal{M}_{\pm 2}(n,\kappa,\mu) - \mathcal{M}_{\pm 1}(n_2,\kappa,\mu) \!=\! 
-2 (n \!-\! n_2) \mp 2 i\!\! \int_{n_2}^{n} \!\! \frac{d n' \mu}{\chi(n')} . 
\qquad
\end{eqnarray}
The small $z(n,\mu)$ expansion of expression (\ref{M1def}) defines the simple 
$\kappa$ dependence of $f_{\pm 1}(n,\kappa,\mu)$; $f_{\pm 2}(n,\kappa,\mu)$ 
depends even more weakly through the lower limit $n_2 \equiv n_{\kappa} + 4$.

Changing variables from $\kappa$ to $n_{\kappa}$ gives,
\begin{equation}
i\Delta_{\pm {\rm IR}}(x;x) = \int_{0}^{n-4} \!\!
\frac{dn_{\kappa} [1 \!-\! \epsilon(n_{\kappa})] \chi^2(n_{\kappa})}{32 \pi^3 G} 
\times \Bigl[ e^{f_{\pm 2}(n,\kappa,\mu)} - e^{f_{\pm 1}(n,\kappa,\mu)} \Bigr] \; .
\label{delta IR final}
\end{equation}
We at length recover the nonlocal contribution to the effective potential by
substituting (\ref{delta IR final}) in expressions (\ref{DVfermion}) and 
(\ref{fermiontrace}) and integrating the derivative,
\begin{equation}
\Delta V'_{\rm IR} = - 4 f^2 \varphi {\rm Re}\Bigl[ i\Delta_{+ {\rm IR}}(x;x)\Bigr]
+ 2 f H \Bigl(2 \!+\! \frac{d}{dn} \Bigr) {\rm Im}\Bigl[i\Delta_{+ {\rm IR}}(x;x)
\Bigr] \; . \label{DVIR}
\end{equation}
The inflaton is assumed constant but expression (\ref{delta IR final}) obviously
depends on the past history of the inflationary geometry. No such term could be
subtracted off by a classical action. Note also that we expect $\Delta V_{\rm IR}$
to be numerically smaller that $\Delta V_{1}$ because it involves only a portion 
of the full Fourier mode sum, and because the integrand is a difference between 
the approximations (\ref{M2def}) and (\ref{M1def}).  

\section{Conclusions}

We have derived an analytic approximation for the contribution of a Yukawa-coupled
fermion (\ref{Lfermion}) to the effective potential of the inflaton in the presence
of a general inflationary background geometry (\ref{geometry}). We start from the
exact expression (\ref{DVfermion}) for the derivative of this potential in terms of
the trace of the coincident limit of fermion propagator with mass $m = f \varphi$
in the presence of a constant inflaton. That coincidence limit is then represented 
(\ref{fermiontrace}) in terms of the coincidence limits of scalar propagators 
$i\Delta[\xi_c,M^2_{\pm}](x;x)$ with conformal coupling $\xi_c = \frac14
(\frac{D-2}{D-1})$ and complex conjugate masses $M^2_{\pm} = m (m \mp i H)$. These
propagators are represented as Fourier mode sums (\ref{scalar propagator}) of
mode functions $u(t,k,M_{\pm})$ whose dimensionless complex amplitude 
$\mathcal{M}_{\pm} \equiv \ln[u(t,k,M_{\pm}) u^*(t,k,M_{\mp})/\sqrt{8\pi G} \,]$ 
obeys equation (\ref{Meqn}) with initial conditions (\ref{initial conditions M}).
Even though each amplitude $\mathcal{M}_{\pm}$ is complex, the combination that
contributes to $\Delta V'$ is real. 

All of the preceding statements are exact; our approximation concerns solutions for 
the complex amplitude $\mathcal{M}_{\pm}$. In the ultraviolet regime of $k/a(t) \gg
H(t)$ we employ (\ref{M1def}). We prove that the deviation (\ref{gexp}) falls off
like $(H a/k)^4$ in the ultraviolet, Figures \ref{M1real-figure} and 
\ref{M1imaginary-figure} demonstrate that this approximation is excellent until well 
after horizon crossing. Some e-foldings after horizon crossing the ultraviolet
approximation (\ref{M1def}) begins to break down, most strongly for the imaginary 
part of $\mathcal{M}_{\pm}$. A second approximation (\ref{M2def}) then becomes
appropriate, and Figures \ref{M2real-figure} and \ref{M2imaginary-figure} demonstrate
that it remains accurate to the end of inflation. In comparing our analytic 
approximations (\ref{M1def}) and (\ref{M2def}) with the numerical solutions for
$\mathcal{M}_{\pm}$ it was of course necessary to assume a specific background 
geometry. For simplicity we took this to be that of the simple quadratic potential
(\ref{Vclass}), however, Figure \ref{Starobinsky2-figure} shows that our 
approximations become even more accurate for a typical plateau model (\ref{StaroU}).

It is also worth noting that the task of approximating conformally coupled scalar 
propagators with complex masses $M^2_{\pm} = m (m \mp i H)$ seems to be considerably 
simpler than that of approximating minimally coupled scalar propagators with purely
real masses.\footnote{The great simplification seems to derive from the complex mass,
rather than from the conformal coupling.} Our conformally coupled, complex mass case 
requires only two phases, and the slope of (the real part of) $\mathcal{M}_{\pm}$ is 
$-2$ for both of them. In contrast, the minimally coupled real case requires three 
phases, with the slope changing from $-2$ to $-3$ and the final phase exhibiting a 
complicated sort of oscillation \cite{Kyriazis:2019xgj}.

Our result for the effective potential consists of a local part
(\ref{one loop correction}), that comes from the ultraviolet approximation 
(\ref{M1def}), and a numerically smaller nonlocal contribution (\ref{DVIR}) 
that descends from the deviation between late time approximation (\ref{M2def})
and the ultraviolet approximation (\ref{M1def}). The local contribution takes
the form,
\begin{equation}
\Delta V_{1} = -\frac{H^4}{8 \pi^2} \Biggl\{ F\Bigl( \frac{f \varphi}{H},
\epsilon\Bigr) + \Bigl[ (2 \!-\! \epsilon) \Bigl( \frac{f \varphi}{H}\Bigr)^2
+ \Bigl( \frac{f \varphi}{H}\Bigr)^4\Bigr] \ln\Bigl[ \frac{(1 \!-\! \epsilon) 
H}{s}\Bigr] \Biggr\} , \label{localA}
\end{equation}
where the function $F(z,\epsilon)$ is,
\begin{eqnarray}
\lefteqn{ F(z,\epsilon) \equiv \Bigl[ 1 - 2 \epsilon - \frac{2 \epsilon'}{1 \!-\! 
\epsilon}\Bigr] z^2 + (1 \!-\! \epsilon)^2 z^2 } \nonumber \\
& & + \int_{0}^{z} \!\!\! dx \Bigl[ 2 x + 2 x^3\Bigr] 
{\rm Re}\Biggl[\psi\Bigl( \frac12 \!+\! \nu(x)\Bigr) + \psi\Bigl( \frac12 \!-\! 
\nu(x)\Bigr) \Biggr] \nonumber \\
& & \hspace{1cm} + \Bigl[ \frac{d}{d n} - 3 \epsilon\Bigr] 
\!\! \int_{0}^{z} \!\!\! dx \Biggl\{ x {\rm Re}\Biggl[
\psi\Bigl( \frac12 \!+\! \nu(x)\Bigr) + \psi\Bigl( \frac12 \!-\! \nu(x)\Bigr) 
\Biggr] \nonumber \\
& & \hspace{4.5cm} - x^2 {\rm Im}\Biggl[\psi\Bigl( \frac12 \!+\! \nu(x)\Bigr) + 
\psi\Bigl( \frac12 \!-\! \nu(x)\Bigr) \Biggr] \Biggr\} . \qquad \label{localB}
\end{eqnarray}
Note that taking $\epsilon = 0$ in the local contribution (\ref{localA}-\ref{localB}) 
recovers the de Sitter limit of Candelas and Raine \cite{Candelas:1975du}. Note
also that our results confirm indirect arguments \cite{Miao:2015oba} about the 
approximate validity of the de Sitter result for general inflationary geometries 
(\ref{geometry}), and about the existence of a part that depends nonlocally on the 
geometry.

The large $\varphi$ (small $H$) expansion (\ref{large field exp}) begins with 
the classic flat space result of Coleman and Weinberg \cite{Coleman:1973jx} and 
then gives a series of corrections which depend more and more strongly on the 
inflationary geometry,
\begin{eqnarray}
\lefteqn{\Delta V_1 = - \frac{H^4}{8 \pi^2} \Biggl\{ \Bigl( \frac{f \varphi}{H}\Bigr)^4
\ln\Bigl( \frac{f \varphi}{s}\Bigr) - \frac14 \Bigl( \frac{f \varphi}{H}\Bigr)^4 
+ (2 \!-\! \epsilon) \Bigl( \frac{f \varphi}{H}\Bigr)^2 \ln\Bigl(\frac{f \varphi}{s}\Bigr)
} \nonumber \\
& & \hspace{-0.5cm} + \Bigl[ \frac12 \!-\! \epsilon \!+\! \frac23 (1 \!-\! \epsilon)^2 
\!-\! \frac{\epsilon'}{1 \!-\! \epsilon}\Bigr] \Bigl(\frac{f \varphi}{H}\Bigr)^2 
+ \Bigl[ \frac12 \!-\! \frac2{15} (1 \!-\! \epsilon)^4\Bigr] \ln\Bigl( 
\frac{f \varphi}{H}\Bigr) + O(1) \Biggr\} . \qquad \label{largeexp}
\end{eqnarray}
The corresponding small field expansion (\ref{small field exp}) should be 
relevant to the end of inflation and the epoch of reheating, during which 
the inflaton passes through zero but the Hubble parameter does not,
\begin{equation}
\Delta V_1 = -\frac{H^4}{8 \pi^2} \Biggl\{ \Biggl[ \Bigl( 1 - \gamma
+ \ln\Bigl[ \frac{(1 \!-\! \epsilon) H}{s}\Bigr]\Bigr) (2 - \epsilon)
- \frac32 \epsilon - \frac{2 \epsilon'}{1 \!-\! \epsilon} \Biggr] \Bigl( 
\frac{f \varphi}{H}\Bigr)^2 + \dots \Biggr\} . \label{smallexp}
\end{equation}

Both of the models we studied begin inflation far into the large field regime.
For the quadratic model (\ref{Vclass}) the initial values are,
\begin{equation}
\phi_0 = 15 \;\; , \;\; \chi_0 \simeq 4.3 \times 10^{-5} \qquad \Longrightarrow
\qquad \frac{\phi_0}{\chi_0} \simeq 3.4 \times 10^5 \; .
\end{equation}
While the plateau model (\ref{StaroU}) has,
\begin{equation}
\phi_0 = 5.3 \;\; , \;\; \chi_0 \simeq 6.4 \times 10^{-6} \qquad \Longrightarrow 
\qquad \frac{\phi_0}{\chi_0} \simeq 8.5 \times 10^5 \; .
\end{equation}
Hence the effective potential is at first essentially the leading term of the 
large field expansion (\ref{largeexp}). In contrast, the classical potential is
about $V \simeq 3 H^2/8\pi G$. Hence the ratio of the magnitude of the effective 
potential to the classical potential is,
\begin{equation}
\frac{\vert \Delta V_1 \vert}{V} \simeq \frac{\chi^2}{24 \pi^2} \times 
\Bigl( \frac{f \phi}{\chi}\Bigr)^4 \ln\Bigl(\frac{f \phi}{\sigma}\Bigr) \; .
\end{equation}
The size of the logarithm depends on the dimensionless renormalization scale 
$\sigma$, but the other factors are initially,
\begin{eqnarray}
{\rm Quadratic} & \Longrightarrow & \frac{\chi_0^2}{24 \pi^2} \times 
\Bigl( \frac{f \phi_0}{\chi_0}\Bigr)^4 \simeq \Bigl(1 \times 10^{11}\Bigr)
\times f^4 \; , \\ 
{\rm Plateau} & \Longrightarrow & \frac{\chi_0^2}{24 \pi^2} \times 
\Bigl( \frac{f \phi_0}{\chi_0}\Bigr)^4 \simeq \Big( 9 \times 10^{12}\Bigr)
\times f^4 \; .
\end{eqnarray}
We therefore conclude that the effective potential will be quite significant
unless the Yukawa coupling $f$ is so small as to make reheating inefficient.
The Appendix explains why the data strongly disfavor small couplings, which
are not even possible for Higgs inflation \cite{Bezrukov:2007ep} whose top 
quark Yukawa coupling is of order one.

\begin{figure}[H]
\centering
\includegraphics[width=7cm]{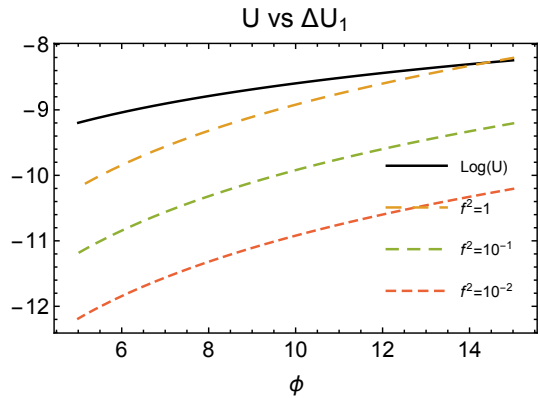}
\caption{\footnotesize{The solid black line is a log-scale plot of
the dimensionless classical potential $(\log_{10} U)$, and the dashed lines 
represent the one loop corrections ($\log_{10}(-\Delta U_1)$) for different 
values of the coupling $f^2$ after subtracting the first three terms in 
equation (\ref{large field exp})}.}
\label{U1subtracted}
\end{figure}

It should also be noted that the first three terms in the large field expansion 
(\ref{largeexp}) can be subtracted off because renormalizability is not
an issue in quantum gravity and we are allowed to change the Lagrangian by any 
function of the inflaton and the Ricci scalar $R = 6 (2 - \epsilon) H^2$. In 
this case, the remaining terms in the series (\ref{largeexp}) represent 
the unavoidable quantum correction $\Delta U_1$. These terms are small for $f^2 
\sim 10^{-6}$, but they can become comparable to the classical potential for 
larger values of the coupling constant. In Figure~\ref{U1subtracted} below we 
plot the one loop potential after the subtraction for different values of $f^2$.

Our results should facilitate precision studies of subtraction schemes
\cite{Liao:2018sci,Miao:2019bnq}, and in the more promising strategy of trying to
cancel the positive contributions to $\Delta V$ from scalars 
\cite{Kyriazis:2019xgj} and photons \cite{Katuwal:2021kry} against the negative 
contributions from fermionic fields that we have derived here. Beyond demonstrating
the potential for such cancellations, a past study was limited by its reliance
on de Sitter results for the effective potentials \cite{Miao:2020zeh}. Now that we
have extended these results to a general inflationary background (\ref{geometry}) 
for minimally coupled scalars \cite{Kyriazis:2019xgj}, for electrically coupled 
photons \cite{Katuwal:2021kry} and for Yukawa coupled fermions (this paper), the
viability of cancellations can be re-examined. We believe that the inclusion of
scalars with arbitrary conformal couplings will provide free parameters that can
be exploited to enforce cancellation to a high order in the large field expansion. 
A potential obstacle is the differing number of derivatives of $\epsilon$ that 
the extended results show: scalars have zero derivatives \cite{Kyriazis:2019xgj},
our work herein has found one derivative for fermions, and there are two derivatives
for photons \cite{Katuwal:2021kry}. It hardly needs to be said that the discovery 
of a viable model would be fascinating owing to the intimate connection between the 
epoch of inflation and the subsequent epoch of reheating.

Finally, it should be emphasized that we have computed the inflaton {\it 
effective potential}, which is defined by setting the inflaton to be a constant. 
This is what one usually wants for studying phase transitions but its suitability 
for inflation might be questioned because the inflaton varies enormously over the
course of inflation. So long as $\epsilon \ll 1$ the inflaton only varies slowly 
and one ought to be able to treat the effective potential as part of the classical 
potential. However, it would be simple enough to extend the approximation scheme
we have developed to a slowly varying inflaton. In particular, every step of the
analysis described in the first paragraph of this Conclusion would apply even for
a time-dependent inflaton. The ultraviolet approximation (\ref{M1def}) ought still 
to apply until well after horizon crossing, so only the nonlocal part might change.

\vskip 1cm

\centerline{\bf Acknowledgements}

This work was partially supported by NSF grants PHY-1806218 and PHY-1912484,
and by the Institute for Fundamental Theory at the University of Florida.

\section{\large Appendix: Connecting Reheating and Fine Tuning}

The universe must reheat before the onset of Big Bang Nucleosynthesis but 
this seeming lower bound can only be achieved through a high degree of fine
tuning. Simple models of inflation all require much higher reheat temperatures. 
Given any model one can use the observed values of the scalar amplitude $A_s$ 
and the scalar spectral index $n_s$ to compute both the number of e-foldings 
from when observable perturbations experienced first horizon crossing to now, 
and also the number of e-foldings from 1st crossing to the end of inflation. 
The difference between these two, $\Delta N = n_0 - n_e$, is the number of 
e-foldings from the end of inflation to now. The reheat temperature $T_R$
can be constrained by comparing a geometrical computation of $\Delta N$ with
a thermal one. 

We follow the geometrical computation by Mielczarek \cite{Mielczarek:2010ag}. 
From the definition of $n$ that the number of e-foldings from any time to the 
present (with $a(t_0) = 1$) is,
\begin{equation}
n \equiv \ln\Bigl[ \frac{a(t)}{a_i}\Bigr] \qquad \Longrightarrow \qquad
n_0 - n = \ln\Bigl[\frac1{a(t)}\Bigr] \; .
\end{equation}
First horizon crossing occurs at $k = H(t_k) a(t_k)$, which means that the
number of e-foldings from first horizon crossing to the present is,
\begin{equation}
n_0 - n_k = \ln\Bigl[ \frac{H(t_k)}{k}\Bigr] \; .
\end{equation}
In the slow roll approximation the scalar power spectrum is,
\begin{equation}
\Delta^2_{\mathcal{R}}(k) \simeq \frac{G H^2(t_k)}{\pi \epsilon(t_k)} \qquad
\Longrightarrow \qquad n_0 - n_k \simeq \frac12 \ln\Bigl[ \Delta^2_{\mathcal{R}}(k)
\!\times\! \frac{\pi \epsilon(t_k)}{G k^2}\Bigr] \; .
\end{equation}
The power spectrum data is well fit using just the scalar amplitude $A_S$,
the scalar spectral index $n_s$ and the pivot wave number $k_0$,
\begin{equation}
\Delta^2_{\mathcal{R}}(k) \simeq A_S \Bigl( \frac{k}{k_0}\Bigr)^{n_s - 1} 
\qquad \Longrightarrow \qquad n_0 - n_{k_0} \simeq \frac12 \ln\Bigl[
\frac{\pi A_S \epsilon(t_{k_0})}{G k_0^2}\Bigr] \; . \label{deln}
\end{equation}
Now compute the number of e-foldings from first horizon crossing to the end of
inflation,
\begin{equation}
\epsilon' \equiv \frac{d\epsilon}{dn} \qquad \Longrightarrow \qquad n_e - n_k
= \int_{\epsilon(t_k)}^1 \frac{d\epsilon}{\epsilon'} \; . \label{deps}
\end{equation}
For the simple quadratic model we studied, the first slow roll parameter obeys,
\begin{equation}
V = \frac12 m^2 \varphi^2 \qquad \Longrightarrow \qquad \epsilon(t_{k_0}) \simeq
\frac14 (1 \!-\! n_s) \;\; , \;\; \epsilon' \simeq 2 \epsilon^2 \; . \label{model} 
\end{equation}
Relations (\ref{model}) are the largest form of model dependence. Combining them 
with (\ref{deln}) and (\ref{deps}) gives the number of e-foldings from the end of 
inflation to the present,
\begin{equation}
\Delta N = \frac12 \ln\Biggl[ \frac{\pi (1 - n_s) A_s}{4 G k_0^2}\Biggr] -
\frac2{1 - n_s} + \frac12 \; . \label{DeltaN1}
\end{equation}
With 2015 Planck numbers \cite{Ade:2015lrj} this works out to about $\Delta N 
\simeq 66.6$.

Now compute $\Delta N$ thermally by following the portion of the inflaton's 
kinetic energy density,
\begin{equation}
\rho_e \equiv \frac12 \dot{\varphi}^2 = \frac{\epsilon H^2}{8\pi G} \simeq
\frac{(1 \!-\! n_s)^2 A_S}{128 G^2} \; , 
\end{equation}
that thermalizes at the end of reheating,
\begin{equation}
\rho_R = \rho_e \Bigl( \frac{a_e}{a_R}\Bigr)^3 = \frac{g_* \pi^2 T_R^4}{30} 
\qquad \Longrightarrow \qquad n_R - n_e \simeq \frac13 \ln\Bigl[
\frac{15 (1 \!-\! n_s)^2 A_S}{64 \pi^2 g_* G^2 T_R^4}\Bigr] \; . \label{stepA}
\end{equation}
Here $g_*$ is the number of relativistic species. At recombination we have,
\begin{equation}
\frac{a_{rec}}{a_R} = \Bigl(\frac{g_*}{2}\Bigr)^{\frac13} \times 
\frac{T_R}{T_{rec}} \qquad \Longrightarrow \qquad n_{rec} - n_R = \frac13
\ln\Big[ \frac{g_* T^3_R}{2 T^3_{rec}}\Bigr] \; . \label{stepB}
\end{equation}
And the number of e-foldings from recombination to the present is,
\begin{equation}
\frac{a_0}{a_{rec}} = \frac{T_{rec}}{T_0} \qquad \Longrightarrow \qquad
n_0 - n_{rec} = \frac13 \ln\Bigl[ \frac{T^3_{rec}}{T^3_0}\Bigr] \; .
\label{stepC}
\end{equation}
Combining (\ref{stepA}), (\ref{stepB}) and (\ref{stepC}) causes $g_*$ to
drop out \cite{Mielczarek:2010ag},
\begin{equation}
\Delta N = \frac13 \ln\Biggl[ \frac{15 (1 - n_s)^2 A_s}{128 \pi^2 G^2 T_{R}
T_0^3}\Biggr] \simeq 62.7 - \frac16 \ln(G T^2_R) \; . \label{DeltaN2}
\end{equation}
Equating (\ref{DeltaN1}) and (\ref{DeltaN2}) implies $T_R \simeq 
10^{14}~{\rm GeV}$!

The reason high reheat temperatures are favored is that extrapolations of the 
simple models which describe the observed power spectrum correspond to small 
values of $\Delta N$, requiring large $T_{R}$. Of course the uncertainties on 
$T_R$ are great owing to the exponential dependence on the factor of 
$\frac2{1 - n_s}$ in (\ref{DeltaN1}), but the preference for large reheat 
temperatures is clear. Considering more general models in the context of WMAP 
data, Martin and Ringeval derived a lower bound of more than $10^4~{\rm GeV}$ 
\cite{Martin:2010kz}. These results can only be evaded by decreasing the 
number of e-foldings between first crossing and the end of inflation. 
Arranging that requires tuning the lower portion of the inflaton potential to 
be vastly steeper than the portion during which observable perturbations 
experienced first crossing. Of course this could be done, but it raises obvious 
questions about why the potential changed form, and why the initial condition 
was such that observable perturbations happened to be generated when the scalar 
was on the flat portion.

\end{document}